\documentclass[%
    aps,
    prd,
    letterpaper,
    superscriptaddress,
    preprintnumbers,
    floatfix,
    twocolumn,
    10pt,
    nofootinbib,
    showpacs,
    hyphens
]{revtex4-1}  
\usepackage{amsmath}
\usepackage{amsfonts}
\usepackage{amssymb}
\usepackage{mathtools}
\usepackage{cases}
\usepackage{lmodern}
\usepackage[T1]{fontenc}
\usepackage{upgreek}
\usepackage{microtype}
\usepackage{graphicx} 
\usepackage{comment}    
\usepackage{dcolumn}   
\usepackage{bm}       
\usepackage[caption=false,justification=justified]{subfig}
\usepackage{ragged2e} 
\DeclareCaptionJustification{justified}{\justifying}
\usepackage{multirow}
\usepackage{slashed}
\usepackage{placeins}
\usepackage{epstopdf}
\usepackage{color}
\usepackage{tikz,pgfplots}
\pgfplotsset{compat=newest}
\usepackage{enumerate,comment}
\usepackage{setspace}
\usepackage{dsfont}
\usepackage{nicefrac}
\usepackage{physics}
\usepackage[normalem]{ulem}
\usepackage{tensor}
\usepackage{tabularx}
\usepackage{listings}
\usepackage{adjustbox}
\usepackage{hyperref}  
\hypersetup{breaklinks=true}
\usepackage[capitalize]{cleveref}

\definecolor{c1}{RGB}{206,0,0}
\definecolor{c2}{RGB}{249,149,0}
\definecolor{c3}{RGB}{153,0,210}
\definecolor{c4}{RGB}{0,109,219}
\definecolor{c5}{RGB}{0,146,146}
\definecolor{c6}{RGB}{255,109,182}
\pgfplotscreateplotcyclelist{cycle1}{
{c1}, {c2}, {c3}, {c4}, {c5}, {c6}
}
\definecolor{light-gray}{gray}{0.95}
\renewcommand{\op}[1]{\hat{#1}}

\renewcommand{\vec}[1]{\boldsymbol{#1}}

\newcommand{\argmin}{\mathrm{argmin}}

\begin{document}
\title{\texorpdfstring{Small-scale Hamiltonian optimization of interpolating operators \\ for Lagrangian lattice quantum field theory}{Small-scale Hamiltonian optimization of interpolating operators for Lagrangian lattice quantum field theory}}
    \author{Artur Avkhadiev}
        \affiliation{
            Center for Theoretical Physics, Massachusetts Institute of Technology, Cambridge, MA 02139, U.S.A.}
        \affiliation{
            DOE Co-Design Center for Quantum Advantage}
    \author{Lena Funcke}
        \affiliation{Transdisciplinary Research Area ``Building Blocks of Matter and Fundamental Interactions'' (TRA Matter) and Helmholtz Institute for Radiation and Nuclear Physics (HISKP), University of Bonn, Nussallee 14-16, 53115 Bonn, Germany}
    \author{Karl Jansen}
        \affiliation{
            Center for Quantum Technology and Applications, Deutsches Elektronen-Synchrotron DESY, Platanenallee 6, 15738 Zeuthen, Germany}
        \affiliation{
            Computation-Based Science and Technology Research Center, the Cyprus Institute, 20 Kavafi Street, 2121 Nicosia, Cyprus}
    \author{Stefan K{\"u}hn}
        \affiliation{
            Center for Quantum Technology and Applications, Deutsches Elektronen-Synchrotron DESY, Platanenallee 6, 15738 Zeuthen, Germany}
    \author{Phiala Shanahan}
        \affiliation{
            Center for Theoretical Physics, Massachusetts Institute of Technology, Cambridge, MA 02139, U.S.A.}
        \affiliation{
            DOE Co-Design Center for Quantum Advantage}
        \affiliation{
            NSF AI Institute for Artificial Intelligence and Fundamental Interactions}
\preprint{
    \vbox{\hbox{{MIT-CTP/{5745}}}}}
\begin{abstract}
Lattice quantum field theory calculations may potentially combine the advantages of Hamiltonian formulations with the scalability and control of conventional Lagrangian frameworks. 
However, such hybrid approaches need to consider (1) the differences in renormalized coupling values between the two formulations, and (2) finite-volume and discretization effects when the Hamiltonian component of the calculation is characterized by a smaller volume or coarser lattice spacing than the Lagrangian component.
This work investigates the role of both factors in the application of Hamiltonian-optimized interpolating operator constructions for the conventional Lagrangian framework.
The numerical investigation is realized for the pseudoscalar meson in the Schwinger model, using tensor-network and Monte-Carlo calculations.
It is demonstrated that tensor-network-optimized constructions are robust to both (1) and (2).
In particular, accurate optimized constructions for the pseudoscalar meson can be obtained from calculations with a smaller number of Hamiltonian lattice sites, even when the meson mass itself receives significant finite-volume corrections.
To the extent that these results generalize to theories with more complicated spectra, the method holds promise for near-term applications in large-scale calculations of lattice quantum field theory. 
\end{abstract}
\maketitle
\section{Introduction}
\label{sec:intro}
    \par Lattice Quantum Field Theory (LQFT) is an approach to regulating ultraviolet divergences in quantum field theories~\cite{Wilson:1974sk, Osterwalder:1977pc}.
    Practically, it has lead to numerical frameworks to study field theories, including in the nonperturbative regime.
    In particular, Monte-Carlo (MC) approaches to calculations in Lagrangian LQFT formulations have been used to address a myriad of problems spanning high-energy and nuclear physics including, for example, precision studies of flavor physics and hadronic structure~\cite{USQCD:2019hyg,Cirigliano:2019jig,USQCD:2019hee,FlavourLatticeAveragingGroupFLAG:2021npn,USQCD:2022mmc,Lin:2017snn,Bazavov:2019lgz,Kronfeld:2019nfb,Amoroso:2022eow}.
    The success of conventional Lagrangian MC techniques is built on the feasibility of accounting for all uncertainties in a systematically improvable way, and scalability to the most computationally demanding systems of interest. 
    Challenges, however, include its susceptibility to sign problems~\cite{Parisi:1983ae,Gibbs:1986ut,Lepage:1989hd}.
    Alternatively, Hamiltonian formulations of LQFT are amenable to approaches based on quantum information science such as quantum-computer (QC) and tensor-network (TN) calculations~\cite{Banuls:2018jag,Banuls:2019rao,Banuls:2019bmf,Haase:2020kaj,Bauer:2022hpo,Meurice:2022xbk,Catterall:2022wjq,DiMeglio:2023nsa,Bauer:2023qgm}, and may offer advantages over conventional Lagrangian formulations\thinspace---\thinspace arising, for example, from the feature of computationally efficient explicit representations of low-lying physical states, and the ability to avoid sign problems, including for calculations of real-time dynamics. 
    Approaches to numerical field theory computations based on quantum information science have so far been most successful for theories in 1+1 spacetime dimensions~\cite{Banuls2018a,Banuls2019,Banuls2020,Funcke2023a}. 
    Scaling these techniques to non-Abelian lattice gauge theories in 3+1 spacetime dimensions is a formidable computational challenge~\cite{Magnifico2021,Felser2019}, and further development is required for applications to high-energy and nuclear physics systems of interest to be tractable.
    \par Hybrid approaches to LQFT, combining the advantages of Hamiltonian methods with the scalability and control of conventional Lagrangian frameworks, are beginning to emerge~\cite{Avkhadiev:2019niu,Harmalkar:2020mpd,Avkhadiev:2022ttx,Clemente:2022cka,Carena:2022hpz,Funcke:2023jbq,Crippa:2024cqr}.
    In the near term, these approaches could leverage small-scale auxiliary Hamiltonian calculations to achieve higher precision in large-scale principally-Lagrangian calculations, while retaining systematic improvability.
    One task that is potentially suited to a hybrid approach in the near term is the Hamiltonian optimization of interpolating operator constructions for  Lagrangian calculations~\cite{Avkhadiev:2019niu,Avkhadiev:2022ttx}.
    In conventional LQFT calculations, these operators approximately map the vacuum state of a theory to a given state of interest, and are used to construct correlation functions from which the physical properties of the state of interest are computed.
    \par Hamiltonian methods for optimizing interpolating operator constructions may provide an advantage over the corresponding techniques in Lagrangian LQFT in that they can operate directly with representations of interpolated states\thinspace---\thinspace quantum states created by acting with interpolating operators on the vacuum\thinspace---\thinspace as opposed to their rates of decay in Euclidean time.
    In the hybrid approach to interpolating operator optimization, interpolated state representations in Hamiltonian calculations are leveraged to parametrically optimize interpolating operator constructions accessible in both formulations.
    The optimized parameters are then used to reproduce the constructions in the Lagrangian MC framework to increase statistical precision of numerical calculations compared to those with unoptimized constructions.
    However, it has not been previously investigated whether the differences between Hamiltonian and Lagrangian formulations would practically affect the accuracy of optimized constructions in such an approach.
    In particular, it has not been characterized how the accuracy of Hamiltonian-optimized constructions in Lagrangian calculations depends on the differences between the formulations in continuous Minkowski and discretized Lagrangian time, and on the reductions in system size, relative to the corresponding large-scale Lagrangian LQFT, that would be practically necessary in near-term, small-scale Hamiltonian calculations.
    \par This work makes the first steps to quantify the effects of differences in LQFT formulations and system sizes used in the Hamiltonian optimization of interpolating operators on the accuracy of optimized constructions in Lagrangian calculations. 
    A numerical investigation is performed in the Schwinger model, with interpolating operator constructions optimized with TNs. 
    Specifically, while the Hamiltonian formulation corresponding to a given Lagrangian formulation requires a continuum temporal limit, which induces differences in renormalized gauge-field couplings and fermion masses between the corresponding formulations, neglecting these differences is not found to have a significant effect on the accuracy of Hamiltonian-optimized constructions in Lagrangian calculations.
    The accuracy is found to be sensitive only to a severalfold coarsening of the Hamiltonian lattice spacing at fixed physical volume, and robust to similar reductions in physical volume at fixed lattice spacing.
    Since the latter determines the number of lattice sites and drives the computational cost of TN calculations, the results provide an encouraging case for the near-term applications of TN-optimized interpolating operators to large-scale Lagrangian LQFT calculations for systems of interest in high-energy and nuclear physics.
    \par The rest of this manuscript is organized as follows.
    \cref{sec:interpolators} discusses how interpolating operators are used in Lagrangian LQFT, defines the interpolating operator optimization, and describes how Lagrangian LQFT calculations with Hamiltonian-optimized interpolating operators may be affected by differences between Hamiltonian and Lagrangian formulations.  
    \cref{sec:schwinger} describes a proof-of-concept investigation in the Schwinger model, and \cref{sec:numerical} details its numerical implementation and presents the results.
    Finally, \cref{sec:outlook} provides a summary and an outlook.
\section{Interpolating operator optimization}
\label{sec:interpolators}
    \par In LQFT calculations with conventional Lagrangian formulations, physical observables can be obtained from the Euclidean-time dependence of correlation functions ~\cite{Parisi:1983ae}.
    For example, in a given LQFT, let $\ket*{\Omega}$ be the vacuum state and $\ket*{E_1}$, a distinct massive ground state: the lowest-energy state for some nonvacuum set of quantum numbers.  
    Then the corresponding energy gap $E_1$\thinspace---\thinspace the difference between the energies of $\ket*{E_1}$ and $\ket*{\Omega}$\thinspace---\thinspace can be obtained from two-point correlation functions between time-separated `interpolating' operators $O^\dagger(0)$ and $O(t)$ 
    mapping $\ket*{\Omega}$ to a superposition of states characterized by the quantum numbers of $\ket*{E_1}$,
    \begin{equation}
    \label{eq:twopt}
    \begin{aligned}
    C(t) 
	&\equiv \left\langle \Omega \right\rvert
		\hat{\mathcal{O}}(t)	\hat{\mathcal{O}}^\dagger(0)
       \left\lvert \Omega \right\rangle \\
    &= \sum_{n \geq 1} \lvert \bra{E_n}\op{\mathcal{O}}^\dagger(0)\ket{\vphantom{E_n}\Omega} \rvert^2 e^{-E_{n}t} + \ldots
    \end{aligned}
    \end{equation}
    Here, $n$ indexes states $\ket*{E_n}$ in the tower of states with the quantum numbers of the interpolating operators, $E_n$ denotes the energy gap between $\ket*{E_n}$ and $\ket*{\Omega}$, and the ellipsis denotes finite-temperature contributions from the finite temporal lattice extent.
    The energy gap $E_1$ can be extracted as the large-time asymptote of a lattice derivative of $C(t)$, for example
    \begin{equation}
    \label{eq:effective-energy}
    a E_{\mathrm{eff}}\left(t+\frac{1}{2}a\right) 
        = \log\frac{ C(t)}{ C(t+a)} \xrightarrow{t \gg 0} a E_{1}\thinspace,
    \end{equation}
    where $a$ is the lattice spacing.
    \par For a `perfect' interpolating operator construction, 
    all terms with $n>1$ in \cref{eq:twopt} would vanish, and $E_\mathrm{eff}(t)$ would equal $E_1$ for all $t>0$ in \cref{eq:effective-energy}. 
    In practice for nonperturbative theories, it is not typically known how to construct perfect operators, and imperfect constructions with the correct set of quantum numbers are used, resulting in nonzero excited state contributions in \cref{eq:twopt,eq:effective-energy}, particularly at early $t$.
    \begin{figure}[t]
            \centering
                \includegraphics[width=0.99\columnwidth]{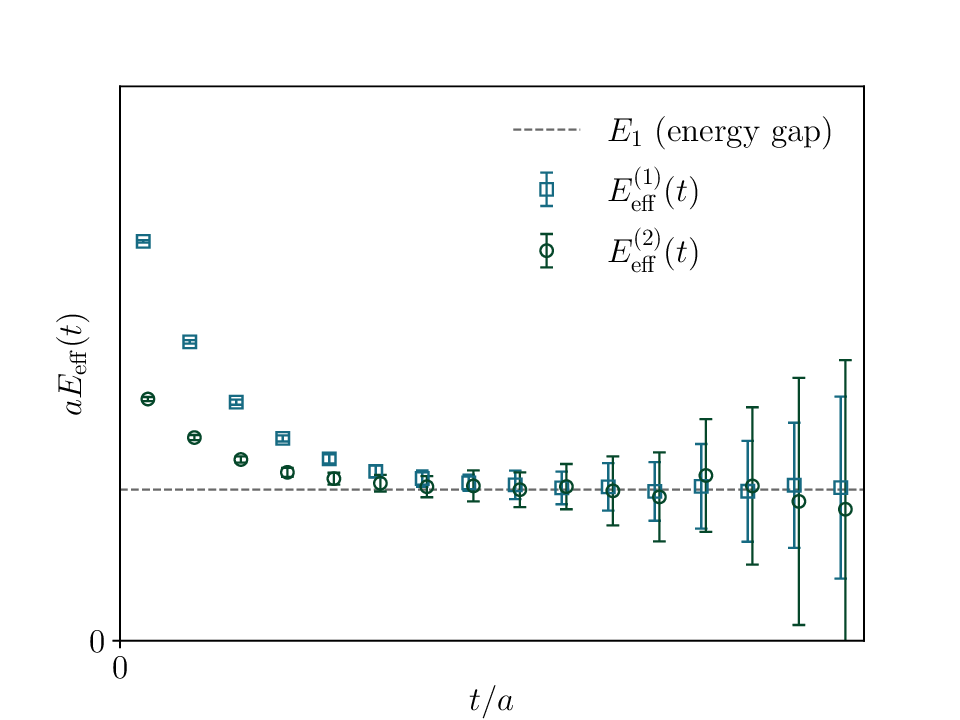}   
                \caption{ 
                A schematic illustration of typical effects of excited-state contributions and decay of StN ratios in conventional LQFT calculations.
                The energy gap $E_1$ can be obtained as the large-$t$ asymptote of statistical fits to stochastic estimates of effective energies $E^{(i)}_{\mathrm{eff}}(t)$ for $i = 1,\,2$, with $i=2$ markers horizontally offset for visual clarity. 
                % Precisely, here $C^{(i)}(t)$ are given by 3-state models (\cref{eq:twopt} for $n = 3$) with $E_1 = 0.56$. \langle C^{(i)}(t) \rangle$ are sampled from log-normal distributions at each $t$, with different variances for $i = 1, 2$ both growing exponentially in $t$. Statistical uncertainties are estimated with $100$ bootstrap resamples.
                \label{fig:effective-mass-sketch}
                }
            \end{figure}
    Furthermore, at late $t$, stochastic estimates of $C(t)$ for calculations of most observables of interest in conventional MC calculations suffer from an exponential decay of signal-to-noise (StN) ratios: an effect known as the StN problem~\cite{Parisi:1983ae, Lepage:1989hd}.
    Typical effects of excited-state contamination and decaying StN ratios on $ E_\mathrm{eff}(t)$ in conventional MC calculations  are illustrated in Fig.~(\ref{fig:effective-mass-sketch}).
    These effects are also relevant in calculating higher-point correlation functions\thinspace---\thinspace cf. Refs.~\cite{Wagman:2024rid,Hackett:2024xnx} for recent developments in addressing the StN problem in LQFT correlations functions.
    \par One way to improve the precision of conventional MC LQFT calculations involving a given state of interest is to reduce excited-state contamination in the expectation value of $C(t)$.
    Within the conventional LQFT approach, this may be achieved using a number of methods such as smearing~\cite{APE:1987ehd,Falcioni:1984ei,Teper:1987wt,Morningstar:2003gk,Hasenfratz:2001hp,Capitani:2006ni,Gusken:1989ad,Gusken:1989qx,Alexandrou:1990dq,Luscher:2010iy,Bali:2016lva}, and variational~\cite{Fox:1981xz, Luscher:1990ck, Michael:1982gb, Michael:1985ne, Blossier:2008tx, Blossier:2009kd, Mahbub:2009nr, Detmold:2014rfa, Detmold:2014hla} and Prony techniques~\cite{Fleming:2004hs, Lin:2007iq, Beane:2009kya, Beane:2009gs}.
    To perform such optimization in a Hamiltonian calculation, an interpolating operator construction may be defined as a linear combination
    \begin{equation}
    \label{eq:linear-combination}
    \mathcal{O}_{\vec{\theta}} = \sum_{i = 1}^{N}\cos(\theta_i)\, \mathcal{O}_i\thinspace,
    \end{equation}
    where $\mathcal{O}_i$ are operators that can be realized in both the Hamiltonian and the conventional Lagrangian calculation, and $\theta_i$ are (generally complex-valued) mixing angles combined in a vector $\vec{\theta}$.
    Given $N$ operators, the construction can be defined with $N-1$ angles, because the overall normalization of the interpolated state is irrelevant to the Euclidean-time dependence of correlation functions such as in \cref{eq:effective-energy}.
    The vector $\vec{\theta}$ may be optimized with methods based on input from Hamiltonian calculations for the same system~\cite{Avkhadiev:2019niu,Avkhadiev:2022ttx}.
    While currently at a proof-of-concept stage, such methods could complement existing conventional approaches.
    In particular, a Hamiltonian optimization of $\vec\theta$ could make use of explicit representations of quantum states, distinguishing it from smearing methods.
    Furthermore, this optimization would be independent of the StN problem affecting variational and Prony techniques. 
    \par The mixing angle $\vec\theta$ parametrizes interpolating operator constructions $\mathcal{O}_{\vec\theta}$ in both the Lagrangian and the Hamiltonian components of the hybrid calculation  according to \cref{eq:linear-combination}. 
    In the Lagrangian formulation, $\mathcal{O}_{\vec\theta}$ appears in correlation functions, such as 
    \begin{equation}
        \label{eq:cfun-theta}
        C(t, \vec\theta) = \langle \Omega \rvert \mathcal{O}_{\vec\theta}(t)	{\mathcal{O}_{\vec\theta}}^{\dagger}(0) \left\lvert \Omega \right\rangle\thinspace.
    \end{equation} 
    The angle $\vec\theta$ affects the magnitude of excited-state contamination in the expectation value of $C(t, \vec\theta)$, and thus the precision with which physical observables can be extracted in conventional Lagrangian LQFT.
    In the hybrid approach, $C(t, \vec\theta)$ is calculated for some optimal value of $\vec\theta$ as determined in a Hamiltonian calculation. 
    A particular choice of objective function for the proof-of-concept demonstration in this work is a ``minimum-energy'' (ME) construction for the mixing angle, denoted $\tilde{\vec{\theta}}_{\mathrm{ME}}$:
    \begin{equation}
    \begin{aligned}
    \label{eq:min-energy}
    \tilde{\vec{\theta}}_{\mathrm{ME}}(\lbrace \tilde{g} \rbrace, \tilde{\Lambda})
        &= \argmin_{{\vec\theta}} 
        \dfrac{\bra*{\widetilde{\Omega}}\mathcal{O}_{{\vec{\theta}}}
            H (\lbrace \tilde{g} \rbrace, \tilde{\Lambda}){\mathcal{O}_{{\vec{\theta}}}}^{\dagger}\ket*{\widetilde\Omega}}{\bra*{\widetilde\Omega}\mathcal{O}_{{\vec\theta}}{\mathcal{O}_{{\vec{\theta}}}}^\dagger\ket*{\widetilde\Omega}}\thinspace. \\
    \end{aligned}
    \end{equation}
    The ME construction is defined to reduce excited-state contamination in the Lagrangian-calculated $C(t, \tilde{\vec{\theta}}_{\mathrm{ME}})$ in the limit of $t \to 0$.
    Here, $\tilde{\Lambda}$ denotes a Minkowski lattice geometry in discretized space and continuous time, distinct from the corresponding Euclidean lattice geometry ${\Lambda}$ in discretized spacetime; $\lbrace \tilde{g} \rbrace$ denotes the set of bare couplings regularized by $\tilde{\Lambda}$, distinct from the corresponding set  $\lbrace g \rbrace$ in the Lagrangian calculation. $H(\lbrace \tilde{g} \rbrace, \tilde{\Lambda})$ denotes the Hamiltonian for the same theory as in the Lagrangian component of the calculation.
    The corresponding vacuum state, denoted by
    $\ket*{\widetilde\Omega}$, is obtained independently of $\ket*{\Omega}$ in the Lagrangian calculation, where the latter state is represented implicitly by a MC ensemble of field configurations.
    If $\ket*{\widetilde{\Omega}} = \ket*{\Omega}$, the construction defined by $\tilde{\vec{\theta}}_{\mathrm{ME}}$ results in a correlation function with the smallest possible amount of excited state contamination as $t \to 0$ for the given set of operators.
    In practice, as described further below, $\ket*{\widetilde{\Omega}} \neq \ket*{\Omega}$; therefore, $\tilde{\vec\theta}_\mathrm{ME}$ may result in correlation functions with excited-state contamination above the theoretical minimum.
    These effects are generally expected in Hamiltonian constructions~\cite{Avkhadiev:2022ttx}.
    \par In practice, the following reasons may lead to deviations between optimal parameters of interpolating operator constructions in Hamiltonian and Lagrangian calculations.
    \begin{itemize}
        \item The first reason stems from the differences between sets of bare regularized couplings $\lbrace \tilde{g} \rbrace$ and $\lbrace g \rbrace$ in Hamiltonian and Lagrangian formulations.
        For a given system, the Hamiltonian LQFT may be obtained from the Lagrangian LQFT in the limit of vanishing lattice spacing in the temporal direction, so that $\lbrace g \rbrace \neq \lbrace \tilde{g} \rbrace$ for the same values of physical parameters due to effects induced by RG flow.
        If, instead of performing a matching calculation, the RG-flow effects are neglected by equating the sets of bare couplings $\lbrace \tilde{g} \rbrace = \lbrace g \rbrace$, then $\ket*{\widetilde{\Omega}}$ and $\ket*{\Omega}$ correspond to systems with different values of physical parameters, which in turn affects the precision of MC calculations with interpolating operator constructions defined by $\tilde{\vec{\theta}}_{\mathrm{ME}}(\lbrace \tilde{g} \rbrace, \tilde{\Lambda})$.
        \item The second reason stems from how a Hamiltonian calculation is implemented.
        In practice, such a calculation may be realized, for example, using a TN or a QC.
        In these cases, $\ket*{\widetilde{\Omega}}$ is an upper bound on the ground state of, at best, the target system, or more likely one related to it by a truncation of the original degrees of freedom.\footnote{TN ans{\"a}tze are systematically improvable. 
        In 1+1 spacetime dimensions, it is typically computationally feasible to obtain ground-state TN approximations that are exact within machine precision~\cite{Schollwoeck2011,Orus2014a}.}
        In particular, even as the capabilities of QC and TN techniques in the future approach those required by high-energy and nuclear physics systems, these calculations will likely initially require truncation by a significant reduction of spatial volume in lattice units relative to those in practical Lagrangian calculations.
        The reduction can be achieved by increasing the lattice spacing or decreasing the lattice volume in physical units.
        Results of calculations based on $\ket*{\widetilde{\Omega}}$ and $\ket*{\Omega}$ will then differ by discretization artifacts or finite volume effects, respectively.
    \end{itemize}
    In hybrid approaches that do not take these differences into account, Hamiltonian and Lagrangian observables, such as energy gaps $\tilde{E}_1(\lbrace \tilde{g} \rbrace, \tilde{\Lambda})$ and $E_1(\lbrace g \rbrace, \Lambda)$, would not be exactly equal.
    In Lagrangian calculations making use of Hamiltonian-optimized parameters, such as calculations of $E_1(\lbrace g \rbrace, \Lambda)$ using $\tilde{\vec\theta}_{\mathrm{ME}}(\lbrace \tilde{g} \rbrace, \tilde{\Lambda})$, not taking these differences into account would lead to additional excite-state contamination in  correlation functions, and thus lower precision of statistical fits.
    However, the significance and relative magnitudes of these effects have not been investigated before.
    This work takes the first steps to quantify them, in the context of a toy model.
\section{The Schwinger model}
\label{sec:schwinger}
    \par As a proof of concept, the dependence of Lagrangian calculations with Hamiltonian-optimized interpolating operator constructions on differences in Lagrangian and Hamiltonian formulations is investigated in the Schwinger model~\cite{Schwinger1962,Coleman1975,Coleman1976}, a (1+1)-dimensional theory of quantum electrodynamics that shares a number of key features with quantum chromodynamics (QCD) and may be readily implemented  with conventional Lagrangian MC as well as Hamiltonian TN methods. 
    The theory is considered with a single flavor of massive Wilson fermions in both calculations.
    The Schwinger model is known to have a $J^{PC} = 0^{--}$ bound ground state~\cite{Coleman1976}, a pseudoscalar meson (which may also be called a vector meson: there is no distinction in 1+1 dimensions).
    The calculation of the meson's mass in the Lagrangian formulation, realized using two-point Euclidean correlation functions of pseudoscalar interpolating operators, is chosen in this study as a benchmark calculation for Hamiltonian interpolating operator optimization.
    \par The Lagrangian formulation of the Schwinger model is contained in the partition function $\mathcal{Z}[g, m, \Lambda]$ defined by the path integral over the gauge and fermion fields, 
    \begin{align}
    \label{eq:schwinger-lattice-action}    
        \mathcal{Z}&[g, m, \Lambda]  \nonumber \\
        &\begin{aligned}
        &= \int \mathcal{D}U
        \mathcal{D}\bar{\psi}
        \mathcal{D}\psi
        \exp\bigg[
            \frac{1}{(ag)^2} 
            \sum_{\mathclap{n\in \Lambda}} 
            \mathrm{Re}\left\lbrack 
                1 - U(n)
            \right\rbrack
             \\
            &\qquad+
            a^2 \sum_{\mathclap{n,\,n^\prime \in \Lambda}} 
	       \bar{\psi}(n)
		      D[U, m](n,n^\prime)
	       \psi(n^\prime)
        \bigg]
        \end{aligned} \\
        \label{eq:schwinger-lattice-action-no-fermions}
        &\begin{aligned}
        &= \int \mathcal{D}U 
        \exp\bigg[\frac{1}{(ag)^2} 
            \sum_{\mathclap{n\in \Lambda}} 
            \mathrm{Re}\left\lbrack 
                1 - U(n)
            \right\rbrack \bigg]
        \\ 
            &\qquad\times
            \mathrm{det}(a^2 D[U,\,m])\thinspace.
        \end{aligned}
    \end{align}
    Here, $a$ denotes the lattice spacing and $\Lambda = \{a n\,\mid\, n = (n_1, n_2) \}$ denotes a Euclidean lattice of physical volume $L \times L_t$, with $1 \leq n_1 \leq L/a$ and $1 \leq n_2 \leq L_t/a \}$.
    The gauge field is described in terms of the gauge group elements $U_{\mu}(n)$, $\mu \in \{ 1,\,2\}$ on links, where
    $U_{\mu}(n)$ denotes a complex-valued link variable directed from site $n$ to site $n + \hat{\mu}$.
    Link variables enter the gauge action in terms of the plaquettes $U(n) \equiv U_{1}(n) U_{2}(n+\hat{1}) U^*_{1}(n + \hat{2}) U^*_{2}(n)$.
    The fermionic field is described in terms of $2$-component spinors $\bar{\psi}(n)$ and $\psi(n)$ on sites $n$ and is characterized by the bare mass $m$ and the bare gauge-field coupling $g$.
    The fermions enter the action through the Wilson Dirac operator,
    \begin{equation}
    \label{eq:wilson-operator-2d}
    \begin{aligned}
        D&[U,\,m](n, n^\prime) \\
            &= 
            % This constant is absorbed in the redefinition of fermion fields
            %\left(m_0 + \frac{2 r}{a}\right)
             \delta(n, n^\prime) - \frac{1}{2(a m + 2)} \\
             &\qquad\quad\;\;\;\times \sum_{\mathclap{\mu \in \{1, 2\}}}\,\,\Big( (1-\gamma^{\mathrm{E}}_{\mu})\,U_{\mu}(n)  \delta(n+\hat{\mu},\,n^\prime) \\
     			&\qquad\qquad\quad+ (1 +\gamma^{\mathrm{E}}_{\mu})\, U^{\dagger}_{\mu}(n-\hat{\mu})\,\delta(n-\hat{\mu}, n^\prime)\Big)\thinspace.
    \end{aligned}
    \end{equation}
    The Clifford algebra of the gamma matrices in Euclidean spacetime is defined by 
    $\lbrace \gamma^{\mathrm{E}}_\mu, \gamma^{\mathrm{E}}_\nu \rbrace = 2 \delta_{\mu\nu}$, where $\delta_{\mu\nu} = \mathrm{diag}(+1, +1)$. 
    Chiral projections require $\gamma_{5} = i \gamma^{\mathrm{E}}_{1} \gamma^{\mathrm{E}}_2$.
    In \cref{eq:schwinger-lattice-action-no-fermions}, $\psi(n)$ and $\bar{\psi}(n)$ are integrated out exactly. 
    \par The Hamiltonian lattice formulation of the model with Wilson fermions reads~\cite{Zache2018,Angelides2023}
    \begin{align}
    \begin{aligned}
        H&[\tilde{g}, \tilde{m}, \tilde{\Lambda}] \\
        &= \sum_n\biggr[\bar{\tilde{\psi}}(n)\left(\frac{1-i\gamma^{\mathrm{M}}_1}{2\tilde{a}}\right)\tilde{U}_1(n)\tilde{\psi}(n+1) + \text{h.c}. \\
        &\quad\qquad+ \left(\tilde{m}-\frac{1}{\tilde{a}}\right)\bar{\tilde{\psi}}(n)\tilde{\psi}(n) + \frac{\tilde{a}\tilde{g}^2}{2}\ell(n)^2 \biggr]\thinspace,
    \end{aligned}
    \label{eq:SchwingerHamiltonian}
    \end{align}
    where $\tilde{a}$ is the lattice spacing on the spatial Minkowski-space lattice $\tilde{\Lambda} =  \lbrace  \tilde{a} n\,\mid\, 1 \leq n \leq \tilde{L}/\tilde{a} \}$ of spatial extent $\tilde{L}$.
    A temporal gauge is fixed, setting temporal link variables $\tilde{U}_0(n) = 1$ on all sites.
    The gauge field is described in terms of spatial link variables $\tilde{U}_1(n)$ and dimensionless electric field variables $\ell(n)$ canonically conjugate to the spatial component of the gauge field, which fulfill the commutation relations $[\ell(n), \tilde{U}_1(n')] = i\delta(n, n^\prime)$.
    The fermionic field is described in terms of $2$-component spinors $\bar{\tilde\psi}(n)$ and $\tilde\psi(n)$ on sites $n$ and is characterized by the bare mass $\tilde{m}$ and the bare gauge-field coupling along the spatial direction $\tilde{g}$.  
    The Clifford algebra of the gamma matrices in Minkowski spacetime is defined by 
    $\lbrace \gamma^{\mathrm{M}}_\mu, \gamma^{\mathrm{M}}_\nu \rbrace = 2 \eta_{\mu\nu}$, where $\eta_{\mu\nu} = \mathrm{diag}(-1, +1)$. 
    Chiral projections require $\gamma_{5} =  \gamma^{\mathrm{M}}_{0} \gamma^{\mathrm{M}}_{1}$.
    The physical states $\ket{\Psi}$ of the Hamiltonian fulfill Gauss's law: $\forall n$ $G(n) \ket{\Psi} = 0$, where $G(n) = \ell(n) - \ell(n-1) - Q(n)$ are the generators for time-independent gauge transformations and  $Q(n) = \psi^\dagger(n)\psi(n) - 1$ is the charge. 
    The values of $\tilde{a}, \tilde{g}$, and $\tilde{m}$ differ from the corresponding $a$, $g$, and $m$ by RG-flow effects induced by taking the limit of the vanishing temporal lattice spacing in the Lagrangian formulation.
    \par The (1+1)-dimensional Schwinger model is superrenormalizable.
    Consequently, in the Hamiltonian formulation with open boundary conditions, gauge fields may be integrated out exactly using Gauss's law.
    Precisely, it is possible to reconstruct the value of the electric field purely from the fermionic charge content according to 
    \begin{equation}
    \label{eq:l-n}
        \ell(n) = \ell_0 + \sum_{k=1}^{n}Q(k)\thinspace,
    \end{equation}
    where $\ell_0$ denotes the electric field value at the corresponding boundary and is set to zero.\footnote{A nonzero background electric field is a physically significant parameter in 1+1 spacetime dimensions and corresponds to the addition of the topological $\theta$-term to \cref{eq:schwinger-lattice-action}~\cite{Coleman:1976uz}.
    This work considers the Schwinger model with $\theta=0$.}
    Integrating out the link variables with \cref{eq:l-n} leads to a more compact description and enables the truncation of the gauge fields to be avoided. % (but is not generalizable to a higher number of spacetime dimensions where there is no superrenormalizability).
    \par The chosen interpolating operator set for the pseudoscalar meson at rest comprises the operators
    \begin{align}
    \label{eq:interp-set-g5-v1}
        \mathcal{O}_{1}&= \sum_n \bar{\psi}(n)\gamma_5\psi(n)\thinspace, \\ 
    \label{eq:interp-set-g5-v3}
        \mathcal{O}_{2}&= \sum_n \bar{\psi}(n) U_{1}(n)\gamma_5\psi(n+\hat{1})\thinspace,
    \end{align}
    where $\psi(n)$, $\bar{\psi}(n)$ and $U_1(n)$ may be defined in either the Lagrangian or the Hamiltonian formulation, and with Euclidean and Minkowski gamma matrices assumed, respectively.
    In either case, the sum is performed at fixed time over all spatial lattice sites.
    For a Hamiltonian formulation with open boundary conditions, if the gauge fields are integrated out exactly, $U_1(n)$ is simply omitted in \cref{eq:interp-set-g5-v3}.
\section{Numerical investigation}
\label{sec:numerical}
    \par Interpolating operator optimization is investigated numerically  in the Schwinger model context described in Sec.~\ref{sec:schwinger}.
    First, interpolating operator constructions are optimized in a Hamiltonian TN calculation.
    Then, the accuracy of Hamiltonian-optimized constructions is studied by applying them in a Lagrangian MC calculation. 
    The restriction to a single flavor of fermions in the theory is made for greater computational efficiency of TN calculations, and Wilson-fermion discretization is chosen to minimize the differences of doubler effects between the formulations~\cite{Susskind:1976jm,Sharatchandra:1981si}.
    Spatial boundary conditions are chosen to be open in the Hamiltonian formulation to allow integrating out the gauge fields, and matching spatial boundary conditions are adopted in the Lagrangian formulation to reduce differences between the spectra.  
    \par The model is characterized by the bare gauge-field coupling, the bare fermion mass, and the physical lattice volume:  
    $a g$, $m/g$, and $Lg \times L_t g$ in the Lagrangian formulation
    and $\tilde{a} \tilde{g}$, $\tilde{m}/\tilde{g}$, and $\tilde{L}\tilde{g}$ in the Hamiltonian formulation.
    Lattice extents in lattice units in the Lagrangian and Hamiltonian formulations are specified by $L/a \times L_t/a$ and $\tilde{L} / \tilde{a}$, respectively. 
    In this work, temporal extents $L_t$ of all lattice volumes in the Lagrangian formulation are chosen such that finite-temperature effects are negligible, and the dependence of Lagrangian calculations on $L_t$ is therefore omitted below.
    The mass of the pseudoscalar meson in the Lagrangian and Hamiltonian formulations is  denoted by $M_\eta(ag, m/g, Lg)$ and $\tilde{M}_\eta(\tilde{a}\tilde{g},\tilde{m}/\tilde{g},\tilde{L}\tilde{g})$, respectively.
    \par With the pseudoscalar interpolating operators $\mathcal{O}_1$ and $\mathcal{O}_2$ defined in \cref{eq:interp-set-g5-v1,eq:interp-set-g5-v3}, respectively, interpolating operator constructions are characterized by a mixing angle $\theta$ according to 
    \begin{equation}
    \label{eq:construction-schwinger}
        \mathcal{O}_\theta = \cos(\theta) \mathcal{O}_1 + \sin(\theta)\mathcal{O}_2\thinspace.
    \end{equation}
    In TN calculations, $O_\theta$ appear in the ME construction in \cref{eq:min-energy} to determine the optimal mixing angle $\tilde{\theta}_\mathrm{ME}(\tilde{a} \tilde{g}, \tilde{m}/\tilde{g}, \tilde{L}\tilde{g})$ for a range of parameters $\tilde{a}\tilde{g}$, $\tilde{m}/{\tilde{g}}$, and $\tilde{L}\tilde{g}$.
    In MC calculations at fixed ${a}{g}$, ${m}/{g}$, and ${L}{g}$, the TN-optimized constructions  appear in correlation functions $C(t, \theta=\tilde{\theta}_\mathrm{ME}(\tilde{a} \tilde{g}, \tilde{m}/\tilde{g}, \tilde{L}\tilde{g}))$ according \cref{eq:cfun-theta}, to quantify the sensitivity of MC calculations to the differences between the formulations.
    \par The following subsections detail how interpolating operator constructions are optimized in the Hamiltonian calculation, how they are applied in the Lagrangian calculation, and how the results of the calculations change as the parameters of the Hamiltonian formulation are varied.
    \subsection{Hamiltonian calculation}
    \par The mixing angles $\tilde\theta_{\mathrm{ME}}(\tilde{a}\tilde{g}, \tilde{m}/\tilde{g}, \tilde{L}\tilde{g})$ and the pseudoscalar meson masses $\tilde{M}_\eta(\tilde{a}\tilde{g}, \tilde{m}/\tilde{g}, \tilde{L}\tilde{g})$ are computed in a Hamiltonian TN calculation for various parameter choices.
    To study the effects of discretization artifacts and RG-induced differences in the gauge coupling, the couplings $\tilde{a}\tilde{g} \in \{ 0.50, 0.67, 1.00, 1.33, 2.00, 4.00 \}$ are considered at fixed  $\tilde{L}\tilde{g} =16$, corresponding to $\tilde{L}/\tilde{a} \in \lbrace 32, 24, 16, 12, 8, 4\rbrace$ lattice sites, and fixed $\tilde{m}/\tilde{g}=0.2$.
    To study the effects of RG-induced differences in fermion masses, choices $\tilde{m}/\tilde{g} \in \lbrace 0.02$, $0.10$, $0.20$, $0.40$, $1.00$, $2.00\}$ are considered at fixed $\tilde{a}\tilde{g}=0.5$ and $\tilde{L}\tilde{g}=16$ ($\tilde{L}/\tilde{a} = 32$).
    Finally, to study the role of finite-volume effects, a set of physical lattice volumes $\tilde{L}\tilde{g} \in \{ 16$, $12$, $8$, $6$, $4$, $2\}$ is used at fixed $\tilde{a}\tilde{g}=0.5$, corresponding  to $\tilde{L}/\tilde{a} \in \lbrace 32, 24, 16, 12, 8, 4\rbrace$, and fixed $\tilde{m}/\tilde{g}=0.2$.
    \par To obtain $\tilde\theta_{\mathrm{ME}}(\tilde{a}\tilde{g}, \tilde{m}/\tilde{g}, \tilde{L}\tilde{g})$ and $\tilde{M}_\eta(\tilde{a}\tilde{g}, \tilde{m}/\tilde{g}, \tilde{L}\tilde{g})$, the ground state $\ket*{\widetilde\Omega}$ and the lowest-energy pseudoscalar state, denoted $\ket*{\tilde{E}_1}$, are computed using Matrix Product States (MPS), a particular kind of one-dimensional TN~\cite{Schollwoeck2011,Orus2014a,Bridgeman2017}. 
    Using the methods from Refs.~\cite{Banuls2013,Angelides2023}, these states are determined variationally for each set of parameters $(\tilde{a} \tilde{g}, \tilde{m}/\tilde{g},  \tilde{L} \tilde{g})$ by minimizing the energy expectation value of the Hamiltonian in \cref{eq:SchwingerHamiltonian} with gauge fields integrated out (see App.~\ref{app:TN} for details). 
    The meson mass $\tilde{M}_\eta(\tilde{a}\tilde{g}, \tilde{m}/\tilde{g}, \tilde{L}\tilde{g})$ then corresponds to the energy difference between the ground state and the excited state upon proper rescaling. 
    For the toy model and small lattice volumes used in this study, the tensor size in the MPS calculations can be increased to a level where numerical errors are comparable to machine precision, so that all TN results of this study are reported without quoting the corresponding uncertainties.\footnote{Typically, a relative error in the energy below $10^{-8}$ is observed for all lattice sizes studied, even for the largest lattice volume with $\tilde{L}/\tilde{a}=32$ sites with bond dimensions of the MPS well below $200$.} 
    \begin{table}[t]
      \centering
      \subfloat[$\tilde{m}/\tilde{g}=0.2$ and $\tilde{L}\tilde{g}=16$ ($8 \leq \tilde{L}/\tilde{a} \leq 32$).\label{tab:configurations-mc}]{
      \renewcommand{\arraystretch}{1.2}
      \begin{tabular}{c|cccccc}
        $\tilde{a}\tilde{g}$
        & $0.50$
        & $0.67$ 
        & $1.00$ 
        & $1.33$
        & $2.00$
        & $4.00$ \\ 
        \hline
        $\frac{1}{\pi}\tilde{\theta}_{\mathrm{ME}}(\tilde{a}\tilde{g})$ 
        & $0.242$ % $0.242038$  
        & $0.235$ % $0.2351865$ 
        & $0.215$ % $0.215178$  
        & $0.188$ % $0.1881395$ 
        & $0.129$ % $0.1284685$
        & $0.040$ % $0.0394693$
        \end{tabular}
      }
      \\
      \subfloat[$\tilde{a}\tilde{g}=0.5$ and $\tilde{L}\tilde{g}=16$ ($\tilde{L}/\tilde{a} = 32$).\label{tab:configurations-mg}]{
      \renewcommand{\arraystretch}{1.2}
        \begin{tabular}{c|cccccc}
        $\tilde{m}/\tilde{g}$
        & $0.02$
        & $0.10$ 
        & $0.20$ 
        & $0.40$ 
        & $1.00$
        & $2.00$ \\
        \hline
        $\frac{1}{\pi}\tilde{\theta}_{\mathrm{ME}}(\tilde{m}/\tilde{g})$ 
        & $0.242$ % $0.2418045$
        & $0.242$ % $0.241929$ 
        & $0.242$ % $0.242038$ 
        & $0.242$ % $0.2421645$
        & $0.242$ % $0.242274$
        & $0.242$ % $0.2422885$
        \end{tabular}
      }
      \\
      \subfloat[ $\tilde{a}\tilde{g}=0.5$ and $\tilde{m}/\tilde{g}=0.2$ ($8 \leq \tilde{L}/\tilde{a} \leq 32$). \label{tab:configurations-lg}]{
      \renewcommand{\arraystretch}{1.2}
        \begin{tabular}{c|cccccc}
        $\tilde{L}\tilde{g}$
        & $16$
        & $12$ 
        & $8$ 
        & $6$ 
        & $4$
        & $2$ \\
        \hline
        $\frac{1}{\pi}\tilde{\theta}_{\mathrm{ME}}(\tilde{L}\tilde{g})$ 
        & $0.242$ % $0.242038$
        & $0.243$ % $0.242658$ 
        & $0.244$ % $0.243933$ 
        & $0.245$ % $0.2452575$ 
        & $0.248$ % $0.248083$
        & $0.259$
        \end{tabular}
      }
      \caption{Hamiltonian-optimized angles $\tilde{\theta}_{\mathrm{ME}}(\tilde{a}\tilde{g}, \tilde{m}/\tilde{g}, \tilde{L}\tilde{g})$, computed as described in the text for various lattice spacings $\tilde{a}\tilde{g}$, fermion masses $\tilde{m}/\tilde{g}$, and physical lattice volumes $\tilde{L}\tilde{g}$ (or the number of lattice sites $\tilde{L}/\tilde{a}$). \\ 
      \label{tab:configurations}
      }
    \end{table}
    \par Given the TN representation of $\ket*{\widetilde{\Omega}}$, optimal mixing angles $\tilde\theta_{\mathrm{ME}}(\tilde{a}\tilde{g}, \tilde{m}/\tilde{g}, \tilde{L}\tilde{g})$, defined by the minimization problem in \cref{eq:min-energy}, are obtained from the solution to a generalized eigenvalue problem specified by the matrix pencil $(H_{ij}, C_{ij})$, where
    \begin{align}
        \label{eq:gevp-h}
        H_{ij} &= \bra*{\widetilde{\Omega}}\mathcal{O}_i H (\tilde{a}\tilde{g}, \tilde{m}/\tilde{g}, \tilde{L}\tilde{g}) {\mathcal{O}_j}^{\dagger}\ket*{\widetilde\Omega}\thinspace,\\
        \label{eq:gevp-c}
        C_{ij} &= \bra*{\widetilde{\Omega}}\mathcal{O}_i  {\mathcal{O}_j}^{\dagger}\ket*{\widetilde\Omega}\thinspace.
    \end{align}
    Here, $H (\tilde{a}\tilde{g}, \tilde{m}/\tilde{g}, \tilde{L}\tilde{g})$ is the Hamiltonian in \cref{eq:SchwingerHamiltonian}, and $i,j = 1,2$ index the chosen basis of interpolating operators defined in \cref{eq:interp-set-g5-v1,eq:interp-set-g5-v3}, respectively, with gauge fields integrated out in all three expressions.
    The operators $\mathcal{O}_i$ with $i = 1,2$ and $H(\tilde{a}\tilde{g}, \tilde{m}/\tilde{g}, \tilde{L}\tilde{g})$ can be expressed as matrix product operators (MPOs). Thus, after obtaining the MPS approximation for the ground state, $\ket*{\widetilde\Omega}$, the matrix elements $H_{ij}$ and $C_{ij}$ can be evaluated by contracting the corresponding TN~\cite{Schollwoeck2011,Orus2014a,Bridgeman2017}.    
    The eigenvectors are parametrized by $(\cos {\theta}, \sin{\theta})$, and ${\theta} = \tilde\theta_{\mathrm{ME}}(\tilde{a}\tilde{g}, \tilde{m}/\tilde{g}, \tilde{g}\tilde{L})$ corresponds to the eigenvector with the smallest eigenvalue. The resulting angles are summarized in \cref{tab:configurations}.
    \subsection{Lagrangian calculation}
    \par The pseudoscalar two-point correlation function $C(t, \theta)$ is
    obtained from a Lagrangian calculation for $(ag, m/g, Lg) = (0.5, 0.2,
    16)$, corresponding to $L/a = 32$ lattice sites in the spatial direction.
    The physical volume of the system is sufficiently large so that
    finite-volume effects in the Lagrangian calculation, as characterized by $e^{- M_\eta L}$ and $e^{- L g}$, are less than $10^{-6}$.
    Since Hamiltonian optimization at $(\tilde{a}\tilde{g}, \tilde{m}/\tilde{g}, \tilde{L}\tilde{g})$ corresponds to the Lagrangian system at $({a}{g}, {m}/{g}, {L}{g})$ up to RG-induced coupling differences, the optimal ME mixing angle in this Lagrangian system is given by $\tilde{\theta}_{\mathrm{ME}}(\tilde{a}\tilde{g},\tilde{m}/\tilde{g}, \tilde{L}\tilde{g})$ with $(\tilde{a}\tilde{g},\tilde{m}/\tilde{g}, \tilde{L}\tilde{g})=(0.5,0.2,16)$ (and $\tilde{L}/\tilde{a} = 32$) in the limit of vanishing differences between the couplings. 
    With nonvanishing differences, $\tilde{\theta}_{\mathrm{ME}}(0.5, 0.2, 16)$ is still expected to correspond to the optimal ME angle most closely out of all the Hamiltonian-calculated angles.
    The angles
    $\tilde{\theta}_{\mathrm{ME}}(\tilde{a}\tilde{g},\tilde{m}/\tilde{g},
    \tilde{L}\tilde{g})$ calculated for other values of Hamiltonian parameters
    are used to characterize the sensitivity of operator optimization to
    discretization artefacts, variations in fermion mass, and finite-volume
    effects in the Hamiltonian formulation.
    \par The correlation function $C(t, \theta)$ is determined stochastically using an ensemble of $N_\mathrm{cfg} = 1000$ gauge field configurations, generated using pseudofermions and the Rational Hybrid MC (RHMC) algorithm to sample according to the probability distribution as defined in \cref{eq:schwinger-lattice-action-no-fermions}, with boundary conditions matching those in the Hamiltonian calculation: open spatial boundary conditions for both gauge and fermion fields, and the usual periodic (antiperiodic) boundary conditions for gauge (fermion) fields in the time dimension.
    A thermalization time of $50$ molecular dynamics (MD) simulation time units is used, as determined by monitoring the plaquette time history.
    The MD time step is set to $0.1$ time units, the length of each MD trajectory to $2.0$ time units, and the MD separation time between saved configurations to $20.0$ time units.
    The ensemble is generated from a single stream. 
    \par Precisely, $C(t, \theta)$ is calculated from a $2\times 2$ matrix of correlation functions $C_{ij}(t)$, $i, j \in \{1, 2\}$ as 
    \begin{equation}
    \begin{aligned}
        \label{eq:c-theta}
        C(t, \theta) &= 
            |\cos (\theta)|^2 C_{11}(t) 
            + \cos(\theta)\sin(\theta)^* C_{12}(t)
             \\
                &\quad+ \cos(\theta)^*\sin(\theta)C_{21}(t)
                + |\sin(\theta)|^2 C_{22}(t)\thinspace,
    \end{aligned}
    \end{equation}
    obtained by inserting \cref{eq:c-theta} into \cref{eq:twopt}.
    Matrix elements $C_{ij}(t) = \bra*{\Omega}\mathcal{O}_i  \mathcal{O}_j^{\dagger}\ket*{\Omega}$ are computed by Wick-contracting the fermion fields into Dirac propagators, evaluated on the MC ensemble as inverses of the Wilson Dirac operator $D[U, m]$ in \cref{eq:wilson-operator-2d}, with the same fermion mass and boundary conditions as those used to generate the configurations.
    To avoid boundary effects arising from the open spatial boundary conditions, only propagators with both source and sink in the bulk of the lattice are used, as defined by the the minimum distance of $L/(4a)$ from the source to either spatial boundary.
    In all measurements, $\mathrm{det}( D[U,m])$ is confirmed to be positive, so the associated re-weighting is unnecessary, 
    while re-weighting to correct for the rational approximation of $D[U,m]$ in RHMC~\cite{Kuberski:2023zky} is neglected.
    \par The correlation function is also calculated for $3$ additional identically generated ensembles of $N_{\mathrm{cfg}}=100$ configurations each, as specified by $ag = 0.5, Lg = 16$  ($L/a=32$), and $m/g \in \lbrace 0.02, 0.1, 0.4 \rbrace$, to compare the scaling of meson masses $M_\eta(m/g)$ and $\tilde{M}_\eta(\tilde{m}/\tilde{g})$ with the respective fermion masses, as a consistency check on the qualitative agreement between Hamiltonian and Lagrangian systems used in this study. 
    This agreement is confirmed as described further in \cref{app:consistency-check}.
    \subsection{Accuracy of Hamiltonian-optimized constructions in Lagrangian calculations
    }
    \begin{figure*}[t!]
        \centering
        \subfloat[Discretization artifacts from varying the lattice spacing $\tilde{a}\tilde{g}$ at fixed physical lattice volume $\tilde{L}\tilde{g}=16$ and fermion mass $\tilde{m}/\tilde{g}=0.2$.  
        \label{fig:contamination_discretization}]{
         \includegraphics[width=0.90\columnwidth]{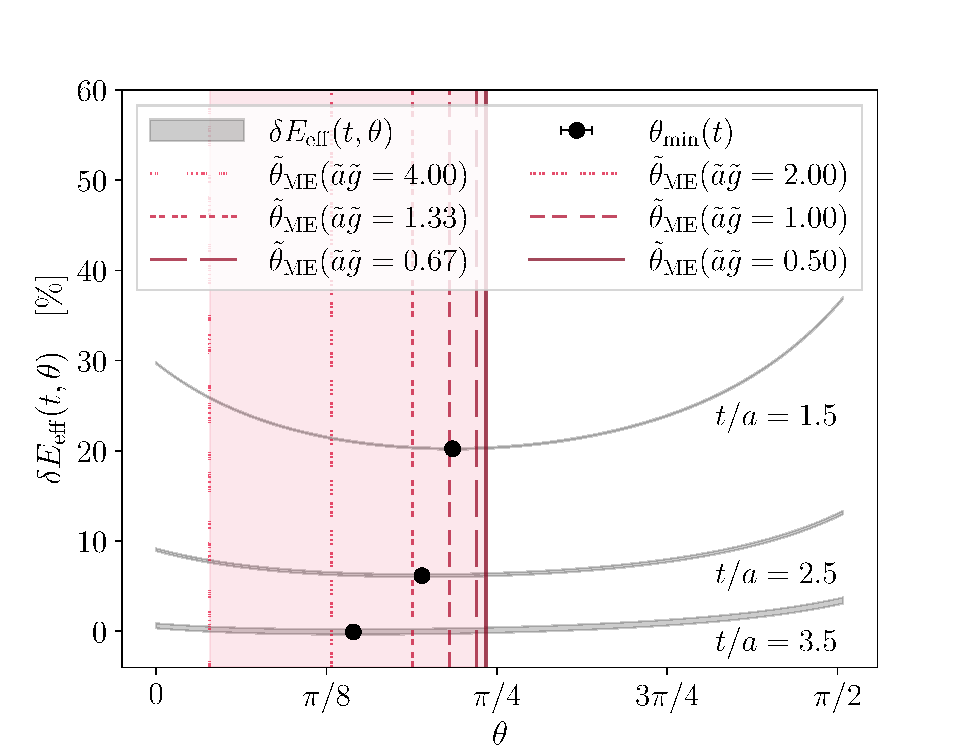}
         \hfill
          \includegraphics[width=0.90\columnwidth]{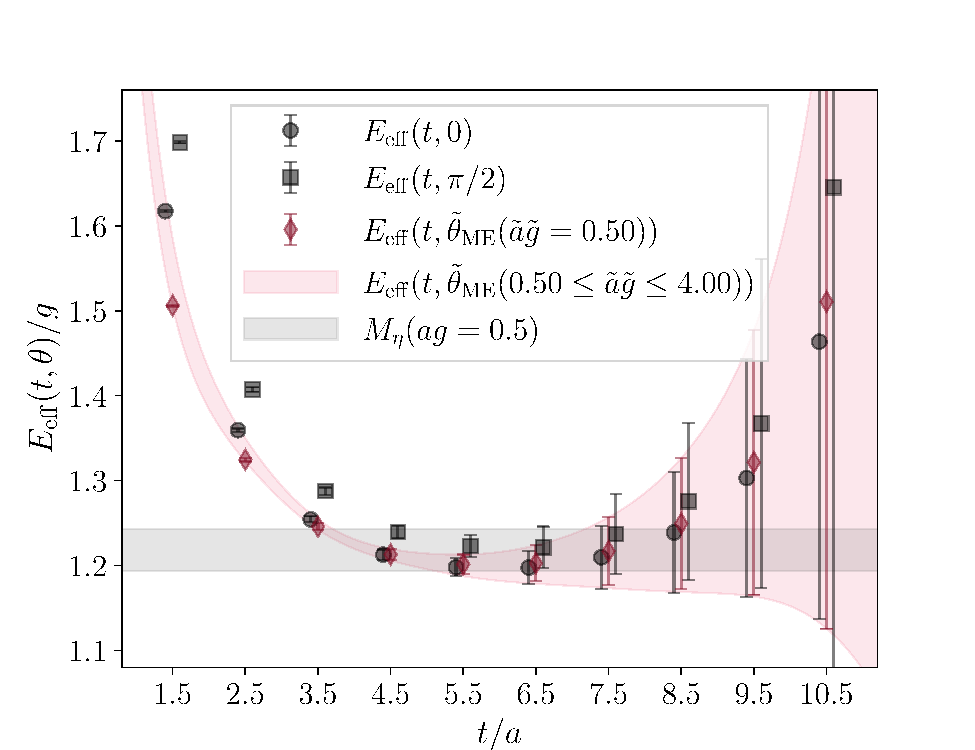}
         }\\[-0.5ex]
        \subfloat[Effects of variations in the fermion mass $\tilde{m}/\tilde{g}$ at fixed lattice spacing $\tilde{a}\tilde{g}=0.5$ and physical lattice volume $\tilde{L}\tilde{g}=16$.
        \label{fig:contamination-mqbare}]{
         \includegraphics[width=0.95\columnwidth]{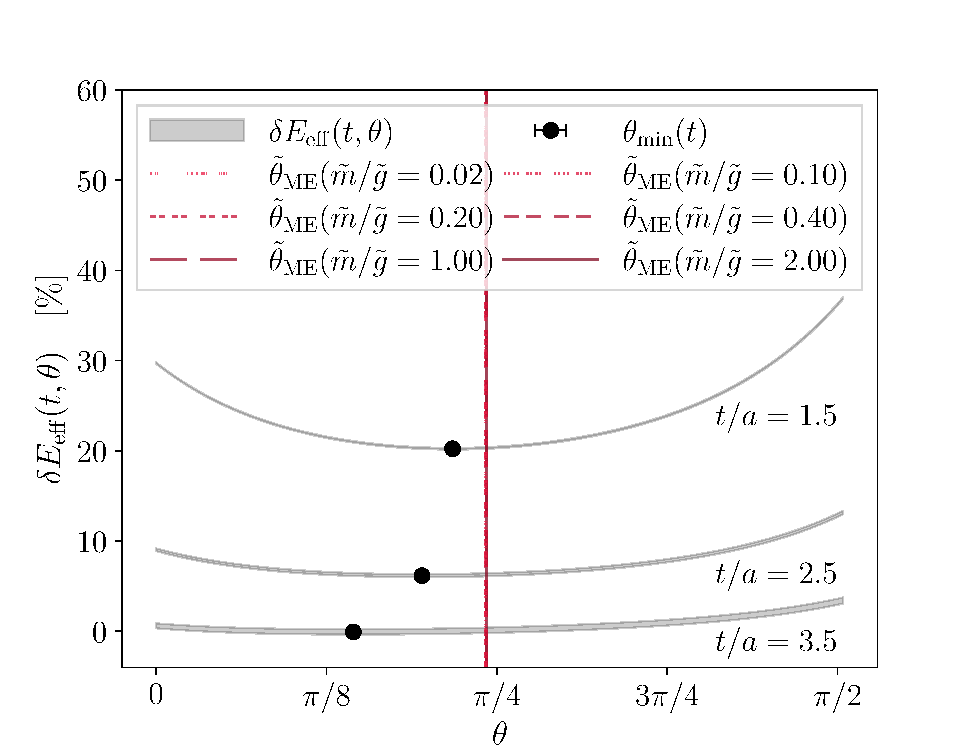}
         \hfill\includegraphics[width=0.95\columnwidth]{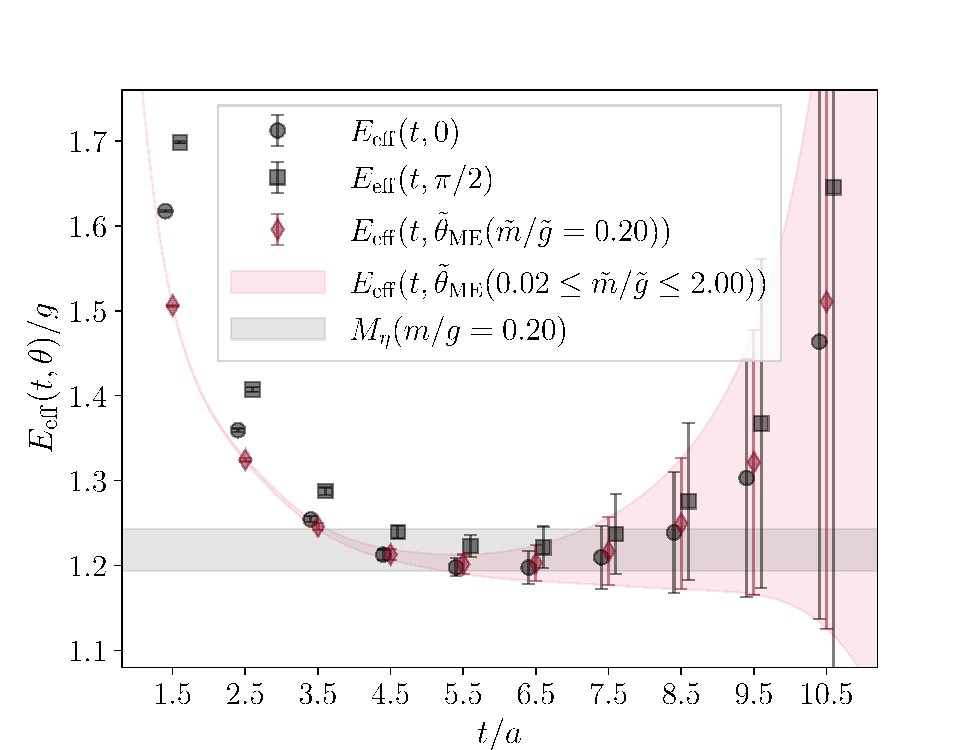}
        }\\[-0.5ex]
        \subfloat[Finite-volume effects from varying the physical lattice volume $\tilde{L}\tilde{g}$ at fixed lattice spacing $\tilde{a}\tilde{g}=0.5$ and fermion mass $\tilde{m}/\tilde{g}=0.2$. The corresponding number of Hamiltonian lattice sites is specified by $\tilde{L}/\tilde{a} = \tilde{L}\tilde{g}/\tilde{a}\tilde{g}$.
        \label{fig:contamination-fve}]{
         \includegraphics[width=0.95\columnwidth]{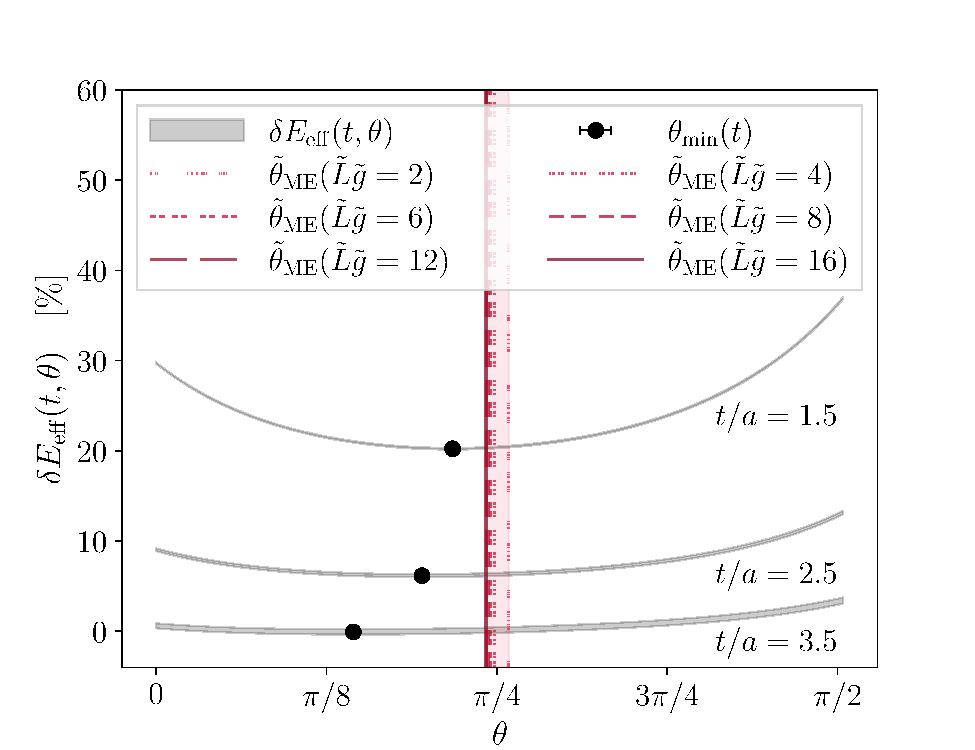}
         \hfill
         \includegraphics[width=0.95\columnwidth]{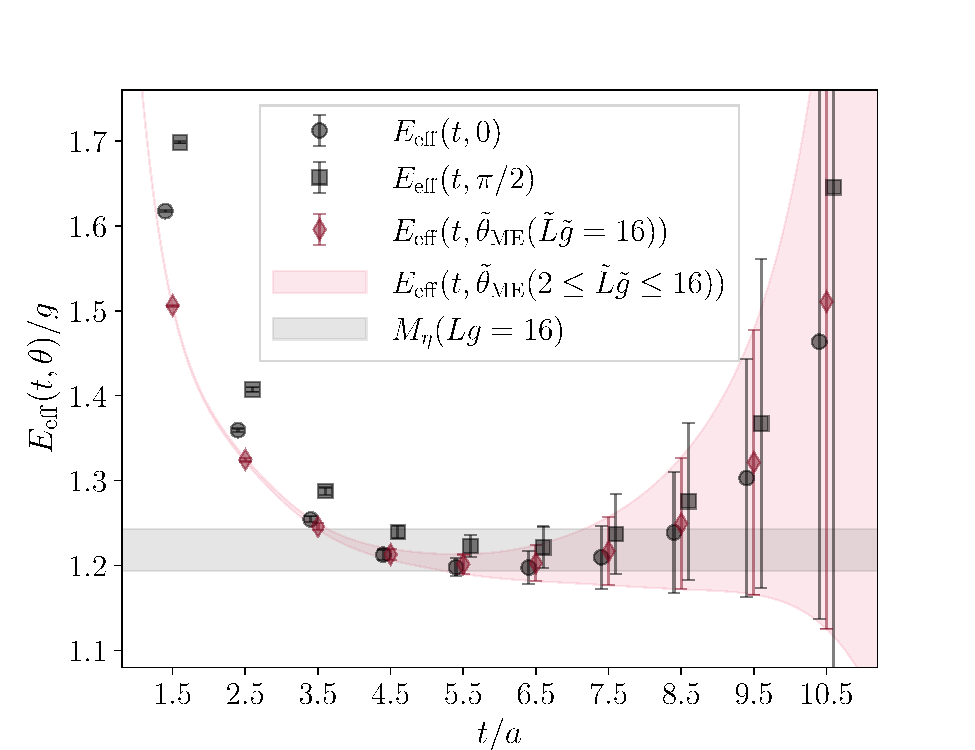}
        }
        \caption{\label{fig:contamination} An illustration of the accuracy of Hamiltonian-optimized mixing angles $\tilde{\theta}_{\mathrm{ME}}(\tilde{a}\tilde{g}, \tilde{m}/\tilde{g}, \tilde{L}\tilde{g})$
        in the Lagrangian calculation of the meson mass $M_\eta$. 
        In each column of figures (left or right), the data is fixed while the pink bands indicate the range of variations in the data as the Hamiltonian parameters $\tilde{a}\tilde{g}$, $\tilde{m}/\tilde{g}$, and $\tilde{L}\tilde{g}$ are varied as described in the captions. 
        \\ \textit{Left panels}: Excited state contamination $\delta E_{\mathrm{eff}}(t,\theta)$ defined in \cref{eq:excited-state-contamination}, the corresponding $\theta$-minima $\theta_{\mathrm{min}}(t)$, and the Hamiltonian-optimized mixing angles $\tilde{\theta}_\mathrm{ME}(\tilde{a}\tilde{g}, \tilde{m}/\tilde{g},\tilde{L}
        \tilde{g})$, with both types of angles computed as described in the text.
        \\ \textit{Right panels}: Effective energies $E_{\mathrm{eff}}(t,\theta)$ for $\theta \in \lbrace 0, \pi/2,  \tilde{\theta}_{\mathrm{ME}}(\tilde{a}\tilde{g}, \tilde{m}/\tilde{g},\tilde{L}
        \tilde{g})\rbrace$ computed according to \cref{eq:effective-energy} and the resulting fit of $M_\eta$ performed as described in the text.
        For visual clarity, markers for $E_{\mathrm{eff}}(t, 0)$ and $E_{\mathrm{eff}}(t, \pi/2)$ are offset horizontally.
        }
    \end{figure*}
    \par To quantify the accuracy of the Hamiltonian-optimized interpolating operator constructions in the Lagrangian calculation, the effective energy $E_\mathrm{eff}(t, \theta)$ is computed from the correlation function $C(t,\theta)$ according to \cref{eq:effective-energy} for all Hamiltonian-optimized $\theta = \tilde{\theta}_{\mathrm{ME}}(\tilde{a}\tilde{g}, \tilde{m}/\tilde{g}, \tilde{L}\tilde{g})$ as well as in a uniform range of $\theta \in [0, \pi/2]$ with spacing $\Delta \theta = 0.01 \times \pi/2$, with $\theta=0$ and $\theta = \pi/2$ corresponding to unoptimized constructions from the local and the nonlocal operators in \cref{eq:interp-set-g5-v1} and \cref{eq:interp-set-g5-v3}, respectively.
    When renormalized couplings between Hamiltonian and Lagrangian formulations are matched and physical volumes are identical, the ME mixing-angle construction $\tilde{\theta}_{\mathrm{ME}}(\tilde{a}\tilde{g}, \tilde{m}/\tilde{g}, \tilde{L}\tilde{g})$, defined in \cref{eq:min-energy}, minimizes excited-state contamination in $E_\mathrm{eff}(t, \theta)$ in the limit $t \to 0$.
    Therefore, the accuracy of operator constructions is characterized by the relative magnitude of excited-state contamination in effective energy, defined as
    \begin{equation}
        \label{eq:excited-state-contamination}
            \delta E_{\mathrm{eff}}(t, \theta) = \frac{E_{\mathrm{eff}}(t,\theta) - M_\eta}{M_\eta}\thinspace,
    \end{equation}
    evaluated at bootstrap level for early Euclidean times $t/a < 4$.
    The effect of variations in $\tilde{a}\tilde{g}, \tilde{m}/\tilde{g}$, and $\tilde{L}\tilde{g}$ on $\tilde{\theta}_{\mathrm{ME}}(\tilde{a}\tilde{g}, \tilde{m}/\tilde{g}, \tilde{L}\tilde{g})$ and $\delta E_{\mathrm{eff}}(t,\tilde{\theta}_{\mathrm{ME}}(\tilde{a}\tilde{g}, \tilde{m}/\tilde{g}, \tilde{L}\tilde{g}))$ is illustrated in the left panel of \cref{fig:contamination}.
    \par The meson mass $M_\eta$ used to compute $\delta E_{\mathrm{eff}}(t, \theta)$ in \cref{eq:excited-state-contamination} is determined by constant fits of $E_{\mathrm{eff}}(t, \theta=0)$ in the plateau region, and is illustrated in the right panel of \cref{fig:contamination}.
    The determined value is $M_\eta/g = 1.22 \pm 0.02$.
    The fits are performed at bootstrap level with $\chi^2$ minimization using the Levenberg-Marquardt algorithm for all fitting windows $[t_{\mathrm{min}}, t_{\mathrm{max}}]$, such that $t_{\mathrm{max}} - t_{\mathrm{min}} \geq 3$, $t_{\mathrm{min}}=2$, and $t_\mathrm{max}$ corresponds to $\mathrm{StN}[E_{\mathrm{eff}}(t_\mathrm{max}, 0)] \geq 5$, where
    \begin{equation}
        \mathrm{StN}[E_{\mathrm{eff}}(t, 0)] = \frac{\bar{E}_{\mathrm{eff}}(t, 0)}{\sigma_{\bar{E}_{\mathrm{eff}}}(t, 0)}\thinspace,
    \end{equation}
    with means $\bar{E}_{\mathrm{eff}}(t, 0)$ and standard errors $\sigma_{\bar{E}_{\mathrm{eff}}}(t, 0)$ estimated from bootstrap resampling.
    To determine the central value and statistical uncertainty in the final determination of $M_\eta$, the results from the largest chosen fit window are used.
    The variations in the central value of $M_\eta$ over the choices of fitting windows are used to determine the systematic uncertainty, which is combined with the statistical uncertainty in quadrature. 
    \par As illustrated in the left panel of \cref{fig:contamination}, the optimized mixing angles $\tilde{\theta}_{\mathrm{ME}}(\tilde{a}\tilde{g}, \tilde{m}/\tilde{g}, \tilde{L}\tilde{g})$ depend most strongly on the Hamiltonian lattice spacing $\tilde{a}\tilde{g}$. This may be explained by the increase in the physical extent of the nonlocal operator $\mathcal{O}_2$, defined in \cref{eq:interp-set-g5-v3}, relative to the physical extent of the meson state as the lattice spacing is increased.
    Consistent with this explanation, $\tilde{\theta}_{\mathrm{ME}}(\tilde{a}\tilde{g}, \tilde{m}/\tilde{g}, \tilde{L}\tilde{g})$  moves closer to $0$ at larger $\tilde{a}\tilde{g}$, which corresponds to a diminishing relative contribution of $\mathcal{O}_2$ to the operator construction in \cref{eq:construction-schwinger}.
    Compared to discretization artifacts, finite-volume effects have a much smaller effect on $\tilde{\theta}_{\mathrm{ME}}(\tilde{a}\tilde{g}, \tilde{m}/\tilde{g}, \tilde{L}\tilde{g})$ when the physical volume $\tilde{L}\tilde{g}$ is reduced at fixed lattice spacing. Finally, at the level of precision in this work, $\tilde{\theta}_{\mathrm{ME}}(\tilde{a}\tilde{g}, \tilde{m}/\tilde{g}, \tilde{L}\tilde{g})$ shows no dependence on fermion mass $\tilde{m}/\tilde{g}$.
    \begin{figure}[t!]
        \centering
        \includegraphics[width=0.98\columnwidth]{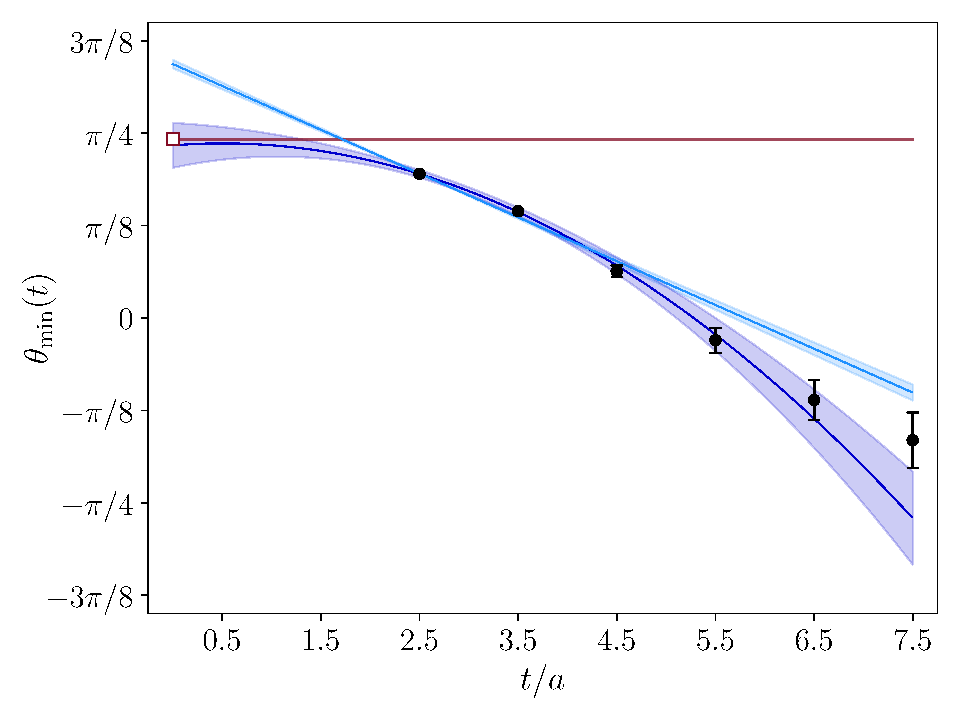}
        \caption{Mixing angles $\theta_{\mathrm{min}}(t)$ minimizing the effective energy $E_{\mathrm{eff}}(t,\theta)$.
        Black circles denote values obtained from Euclidean results, and light (dark) blue bands correspond to linear (quadratic) extrapolations of the results to $t=0$, performed as described in text.
        The horizontal red line with a square marker at $t=0$ corresponds to the Hamiltonian-optimized value $\tilde{\theta}_{\mathrm{ME}}(\tilde{a}\tilde{g}, \tilde{m}/\tilde{g},  \tilde{L}\tilde{g})$ with bare parameters most closely matching those in the Lagrangian formulation.\label{fig:theta}
        }
    \end{figure}
    \par Since  $\tilde{\theta}_{\mathrm{ME}}(\tilde{a}\tilde{g}, \tilde{m}/\tilde{g}, \tilde{L}\tilde{g})$ are defined as minimizing the energy expectation value of the interpolated state as $t \to 0$, they differ from the $\theta$-minima of the Lagrangian-calculated $E_{\mathrm{eff}}(t,\theta)$ at discrete timesteps, which are denoted $\theta_{\mathrm{min}}(t)$ and determined numerically using the Nelder-Mead algorithm, with means and standard errors estimated using bootstrap resampling.
    To confirm that $\tilde{\theta}_{\mathrm{ME}}(\tilde{a}\tilde{g}, \tilde{m}/\tilde{g}, \tilde{L}\tilde{g})$ are consistent with $\theta_{\mathrm{min}}(t)$ for $t \to 0$, linear and quadratic extrapolations of the results are performed.
    Numerical results at $t/a \in \lbrace 0.5, 1.5\rbrace$ are excluded from the fit range, as two-point correlation functions for $t/a \in \{0, 1\}$ receive additional contributions from contact terms in the Wilson Dirac operator in \cref{eq:wilson-operator-2d}.
    The largest time included in the fit range for extrapolations is $t/a=6.5$, with the systematic uncertainty estimated as the variation of the fit parameters when the fit range is increased or decreased by one time step, respectively, and added in quadrature with statistical uncertainties from the original fits.
    The resulting extrapolations are illustrated in \cref{fig:theta}.
    The quadratic extrapolation is consistent with $\tilde{\theta}_{\mathrm{ME}}(0.5, 0.2, 16)$ and, therefore, with negligible RG-flow effects between the  Hamiltonian system at  $(\tilde{a}\tilde{g}, \tilde{m}/\tilde{g}, \tilde{L}\tilde{g}) = (0.5, 0.2, 16)$ and the Lagrangian system at $({a}{g}, {m}/{g}, {L}{g}) = (0.5, 0.2, 16)$.
    As the Hamiltonian parameters are varied, the discrepancy of $\tilde{\theta}_{\mathrm{ME}}(\tilde{a}\tilde{g}, \tilde{m}/\tilde{g}, \tilde{L}\tilde{g})$ with the quadratic extrapolation of $\theta_{\mathrm{min}}(t)$ to $t=0$ is significant only for changes in $\tilde{a}\tilde{g}$ by more than about a factor of $3\text{--}4$ (i.e., when $\tilde{a}\tilde{g}$ is increased from $0.5$ to $1.33\text{--}2.00$ and larger).
    \par For the proof-of-concept application of the optimized operator constructions defined by $\tilde{\theta}_{\mathrm{ME}}(\tilde{a}\tilde{g}, \tilde{m}/\tilde{g}, \tilde{L}\tilde{g})$ in the Lagrangian calculation, the effect of variations in the ME mixing angles is additionally illustrated in the right panel of \cref{fig:contamination}.
    In particular, Hamiltonian optimization for $(\tilde{a}\tilde{g}, \tilde{m}/\tilde{g}, \tilde{L}\tilde{g}) = (0.5, 0.2, 16)$, corresponding to the Lagrangian system at $({a}{g}, {m}/{g}, {L}{g}) = (0.5, 0.2, 16)$ up to RG-flow effects, leads to a reduction of excited-state contamination at $t/a\lesssim 5.5$ relative to unoptimized constructions.
    At the early time $t/a = 1.5$, this reduction is about $30\%$.
    At later times, as expected, the differences in effective energies from different constructions are suppressed by Euclidean time evolution. 
    In particular, the variations in $\theta_{\mathrm{min}}(t)$ for $t/a \gtrsim 4.5$, illustrated in \cref{fig:theta}, lead to variations in $E_{\mathrm{eff}}(t, \theta_{\min})$ negligible within machine precision.  
    Most importantly, the early-time reduction in excited state contamination from Hamiltonian-optimized constructions is robust to changes in the fermion mass by an order of magnitude, and reductions in physical volume and increases in lattice spacing by factors of two.
    Within the precision of this calculation, a change in $E_{\mathrm{eff}}(t, \tilde{\theta}_{\mathrm{ME}})$ is only observable when $\tilde{a}\tilde{g}$ is increased by a factor of $3$--$4$.
    That is, in this application and level of precision one may neglect not only the percent-level differences between $(ag, m/g, Lg)$ and $(\tilde{a}\tilde{g}, \tilde{m}/\tilde{g}, \tilde{L}\tilde{g})$ expected for physically-matched Lagrangian and Hamiltonian calculations~\cite{Kogut:1979vg,Dashen:1980vm,Hasenfratz:1981tw,Karsch:1982ve,Loan:2002ej,Byrnes:2003gg,Funcke:2022opx} but also these much larger differences in the Hamiltonian parameters, with no practical effect on the performance of Hamiltonian-optimized constructions in the Lagrangian calculation.
    \begin{figure}[t!]
        \centering
        \subfloat[
        Discretization artifacts from varying the lattice spacing  $\tilde{a}\tilde{g}$ at fixed physical lattice volume $\tilde{L}\tilde{g}=16$ and fermion mass $\tilde{m}/\tilde{g}=0.2$.\label{fig:convergence-x}]{\includegraphics[width=0.99\columnwidth]{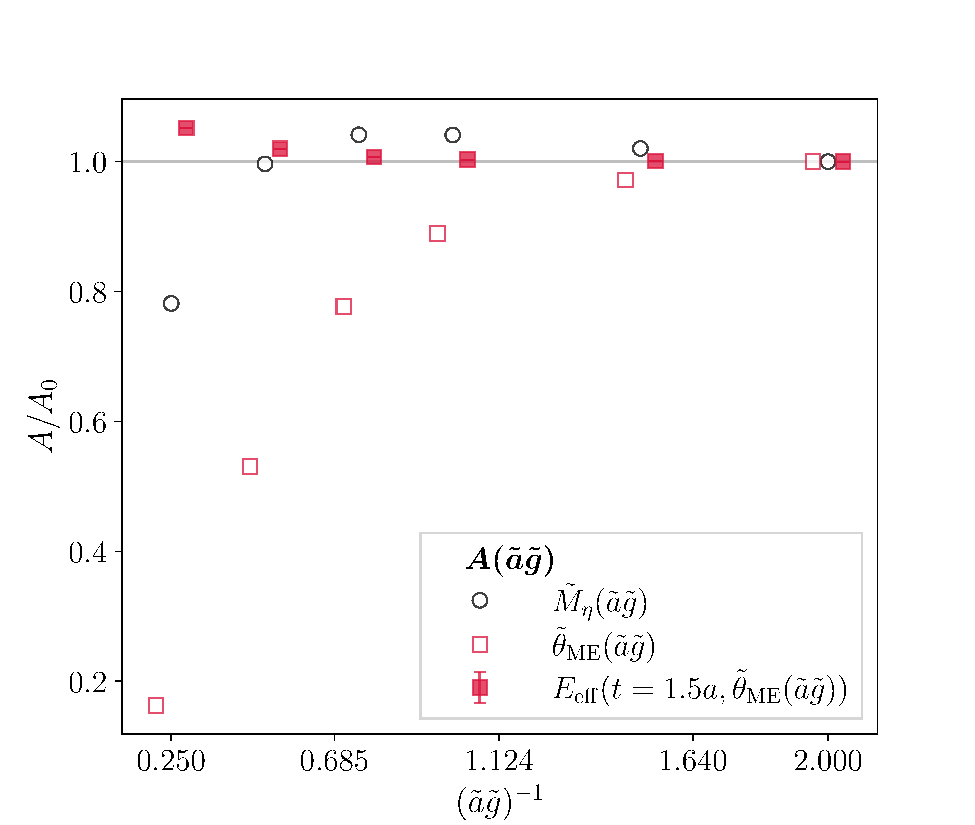}
         }
         \quad
         \subfloat[Finite-volume effects from varying the physical lattice volume $\tilde{L}\tilde{g}$ at fixed lattice spacing $\tilde{a}\tilde{g}=0.5$ and fermion mass $\tilde{m}/\tilde{g}=0.2$. The corresponding number of Hamiltonian lattice sites is specified by $\tilde{L}/\tilde{a} = \tilde{L}\tilde{g} / \tilde{a}\tilde{g}$.
        \label{fig:convergence-N}]{
         \includegraphics[width=0.99\columnwidth]{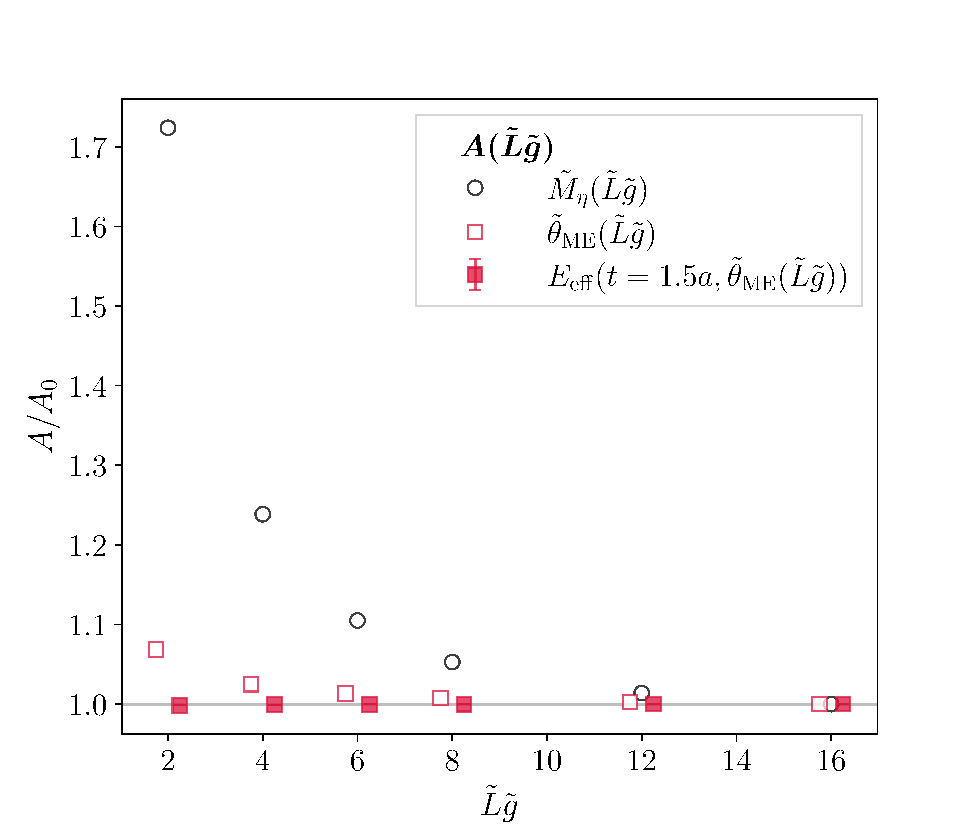}
         }
         \caption{\label{fig:convergence} A comparison of how several observables\thinspace---\thinspace meson mass $\tilde{M}_\eta$, optimal mixing angle $\tilde{\theta}_{\mathrm{ME}}$, and effective energy $E_{\mathrm{eff}}(t, \tilde{\theta}_{\mathrm{ME}})$\thinspace---\thinspace denoted generally $A$, vary as functions of Hamiltonian parameters, and of how they converge to their respective values, denoted $A_0$, 
         at $(\tilde{a}\tilde{g}, \tilde{m}/\tilde{g}, \tilde{L}\tilde{g}) = (0.5, 0.2, 16)$, 
         corresponding to the Lagrangian system at $({a}{g}, {m}/{g}, {L}{g}) = (0.5, 0.2, 16)$ up to RG-flow effects.
         The values $A_0$ are indicated by horizontal lines.
         For visual clarity, markers for $\tilde{\theta}_{\mathrm{ME}}$ and $E_\mathrm{eff}(t, \tilde{\theta}_{\mathrm{ME}})$ are offset horizontally. 
         }
    \end{figure}
    \par The sensitivities of mixing angles $\tilde{\theta}_{\mathrm{ME}}(\tilde{a}\tilde{g}, \tilde{m}/\tilde{g}, \tilde{L}\tilde{g})$ and the effective energy $E_{\mathrm{eff}}(t, \tilde{\theta}_{\mathrm{ME}}(\tilde{a}\tilde{g}, \tilde{m}/\tilde{g}, \tilde{L}\tilde{g}))$ to discretization artifacts and finite volume effects in the Hamiltonian calculation may be compared to each other and to that of the meson masses $\tilde{M}_{\eta}(\tilde{a}\tilde{g}, \tilde{m}/\tilde{g}, \tilde{L}\tilde{g})$ computed directly with TNs, by investigating how these observables converge in $\tilde{a}\tilde{g}$ and $\tilde{L}\tilde{g}$ to their respective values at $(\tilde{a}\tilde{g}, \tilde{m}/\tilde{g}, \tilde{L}\tilde{g})$ = $(0.5, 0.2, 16)$. 
    This comparison is illustrated in \cref{fig:convergence}, using effective energy evaluated at $t/a=1.5$, where it is most sensitive to variations in Hamiltonian parameters.
    The results are consistent with
    $\tilde{\theta}_{\mathrm{ME}}(\tilde{a}\tilde{g}, \tilde{m}/\tilde{g}, \tilde{L}\tilde{g})$
    being respectively more sensitive to discretization artifacts and less sensitive to finite-volume effects than $\tilde{M}_{\eta}(\tilde{a}\tilde{g}, \tilde{m}/\tilde{g}, \tilde{L}\tilde{g})$.
    In particular, there is only a subpercent-level difference between the Hamiltonian-optimized mixing angle $\tilde{\theta}_{\mathrm{ME}}(0.5, 0.2, 8)$, determined at the physical volume $\tilde{L}\tilde{g}=8$ ($\tilde{L}/\tilde{a}=16$ lattice sites), and the best estimate at twice the lattice volume, $\tilde{\theta}_{\mathrm{ME}}(0.5, 0.2, 16)$.
    In contrast, the corresponding difference between TN-calculated meson masses $\tilde{M}_\eta(0.5, 0.2, 8)$ and $\tilde{M}_\eta(0.5, 0.2, 16)$ is more than $5\%$. 
    In summary, even when the TN-calculated meson mass itself receives significant finite-volume corrections, both the TN-optimized constructions for the meson and the resulting improvements in Lagrangian calculations remain robust to significant reductions in physical volume at fixed lattice spacing.
\section{Summary and outlook}
\label{sec:outlook}
    \par This work investigates the feasibility of a hybrid approach to LQFT calculations, where small-scale Hamiltonian optimization of interpolating operator constructions are applied in larger-scale MC calculations for the same theory in a conventional Lagrangian formulation.
    A proof-of-principle numerical investigation is performed for the pseudoscalar meson in the Schwinger model, where the Hamiltonian calculation is implemented with TNs.
    The results indicate that TN-optimized constructions and the resulting improvements in Lagrangian calculations are robust to a significant variation of bare couplings and masses in the Hamiltonian system.
    Specifically, TN-optimized parameters in operator constructions are found to be insensitive to the percent-level differences expected in the bare couplings and the masses between the corresponding Hamiltonian and Lagrangian calculations~\cite{Kogut:1979vg,Dashen:1980vm,Hasenfratz:1981tw,Karsch:1982ve,Loan:2002ej,Byrnes:2003gg,Funcke:2022opx}.
    When coarsening the bare Hamiltonian lattice spacing by a factor of $2$ at fixed physical volume, the TN parameters vary at percent level, and only coarsening by more than a factor of $3$--$4$ leads to observable effects on the suppression of excited-state contamination in Lagrangian calculations at early Euclidean time.
    Furthermore, the optimized parameters remain unchanged as the fermion mass in the Hamiltonian calculation is increased or decreased by an order of magnitude, and
    vary weakly when the physical volume is reduced severalfold at fixed lattice spacing, with subpercent-level variations when the lattice extent is halved to just $16$ sites.
    \par Taken together, the relative robustness of optimized mixing angles and
    the corresponding effective energies to finite-volume effects provide an
    encouraging case for small-scale TN-based optimization of interpolating
    operators, since the computational cost of TN calculations scales primarily
    with the number of lattice sites.
    It is clear, however, that in more complicated systems $E_{\mathrm{eff}}(t, \vec{\theta})$ may depend more sensitively on $\vec{\theta}$, and discretization and finite-volume effects in interpolating operator optimization will become increasingly relevant.
    On the other hand, in theories with more complicated spectra, interpolating operator optimization is expected to result in more significant improvements to Lagrangian calculations\thinspace---\thinspace compared to the $30\%$ reduction in  excited-state contamination at early time already observed in the two-dimensional, Abelian theory used in this work. 
    A way forward for such theories suggested by the results of this study is to require matching of the lattice spacing, but allow a relative reduction of physical volume in the Hamiltonian formulation.
    \par  Future extensions of this approach could investigate alternatives to the ME objective function in \cref{eq:min-energy}, and incorporate estimates of the expected variance of MC estimates of two-point correlation functions, determined by two- and four-point correlation functions of interpolating operators~\cite{Detmold:2014rfa}.
    The definition of an optimized construction could be expanded beyond a linear combination of a pre-determined set of operators in \cref{eq:linear-combination}, and may include constructions with parametrized smearings of fermion and gauge fields. 
    In the Lagrangian component of the calculation, to establish whether the method offers any computational advantage over conventional optimization strategies that rely on MC results and are sensitive to the StN problem, future comparisons should be made in settings where the StN problem is more significant, such as in highly boosted states~\cite{DellaMorte:2012xc,Bali:2016lva,Wu:2018tvt}.
    \par With further development, Hamiltonian interpolating operator optimization may ultimately be used to accelerate large-scale Lagrangian calculations in lattice QCD.
    While standalone applications of methods based on quantum information science, including QC and TN calculations, to non-Abelian gauge theories in 3+1 spacetime dimensions remain a long-term prospect, their hybrid applications with conventional Lagrangian LQFT, such as interpolating operator optimization, may be possible in the nearer term.
    In particular, as state-of-the-art TN calculations have now reached quantum electrodynamics in 3+1 spacetime dimensions, with spatial lattice dimensions of up to $8^3$~\cite{Magnifico2021}, small-scale TN representations of quantum states in non-Abelian gauge theories are within reach.
    The results of this study suggest that the expected systematic uncertainties in such small-scale calculations, including uncertainties from finite-volume effects and those associated with the expressivity of TN ans{\"a}tze, may not be a barrier to the use of TNs for interpolating operator optimization.
    In this way, TNs could be used in the near term to accelerate systematically improvable Lagrangian LQFT calculations at a large scale.
\begin{acknowledgments}
    We thank Tobias Hartung, Joshua Lin, and Ross Young for helpful discussions.
    % Funding
    This work is supported in part by the U.S.~Department of Energy, Office of Science, Office of Nuclear Physics, under grant Contract Number DE-SC0011090 and by the National Quantum Information Science Research Centers, Co-design Center for Quantum Advantage (C2QA) under contract number DE-SC0012704. PES is also supported in part by Early Career Award DE-SC0021006, the Simons Foundation grant 994314 (Simons Collaboration on Confinement and QCD Strings), by the U.S. Department of Energy SciDAC5 award DE-SC0023116, and by the National Science Foundation under Cooperative Agreement PHY-2019786 (The NSF AI Institute for Artificial Intelligence and Fundamental Interactions, http://iaifi.org/).
    This project is supported by the Deutsche Forschungsgemeinschaft (DFG, German Research Foundation) as part of the CRC 1639 NuMeriQS -- project no.\ 511713970. This work is funded by the European Union’s Horizon Europe Framework Programme (HORIZON) under the ERA Chair scheme with grant agreement no.\ 101087126.
    This work is supported with funds from the Ministry of Science, Research and Culture of the State of Brandenburg within the Centre for Quantum Technology and Applications (CQTA). 
    \begin{center}
        \includegraphics[width = 0.08\textwidth]{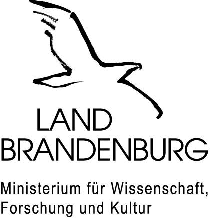}
    \end{center}
\end{acknowledgments}
\appendix
\section{Further details of TN calculations \label{app:TN}}
\par TN states are a family of ans{\"a}tze for the wave function of strongly-correlated quantum many-body systems, based on the entanglement structure of the state. 
The main idea is to break the exponentially large tensor $c_{i_1i_2\dots i_N}$ of the generic wave function
\begin{align}
    \ket{\psi} = \sum_{i_1,i_2,\dots,i_N=1}^d c_{i_1i_2\dots i_N}\ket{i_1}\otimes\ket{i_2} \otimes \dots\otimes \ket{i_N}
    \label{eq:wfct}
\end{align}
for a lattice system with $N$ sites with local bases $\ket{i_k}$, $k=1,\dots,d$, into smaller pieces that can be efficiently stored on a classical computer.\footnote{Note that the local bases for the individual sites do not necessarily need to have the same dimension for a TN approach. 
For notational simplicity, it is assumed here that all dimensions are equal.} Such a decomposition is motivated by the fact that most physically relevant states are very nongeneric, and typically are only moderately entangled~\cite{Hastings2007,Eisert2010}. 
A paradigmatic example is the  Matrix Product States (MPS) ansatz, a particular kind of TN for one-dimensional systems, used in this work. For a system with open boundary conditions, the ansatz reads
\begin{equation}
\begin{aligned}
    \ket{\psi}  &= \sum_{i_1,i_2,\dots,i_N=1}^d M_1^{i_1}M_2^{i_2}\dots M_N^{i_N} \\
    &\quad\qquad\qquad\qquad\times \ket{i_1}\otimes\ket{i_2} \otimes \dots\otimes \ket{i_N}\thinspace,
    \label{eq:MPS}
\end{aligned}
\end{equation}
where the $M_k^{i_k}$ are complex $\chi \times \chi $ matrices for $1<k<N$, and $M_1^{i_1}$ ($M_N^{i_N}$) is a $\chi $-dimensional row (column) vector. 
The parameter $\chi$, called the bond dimension of the MPS, determines the number of variational parameters in the ansatz and limits the amount of entanglement that can be present in the state~\cite{Schollwoeck2011,Orus2014a}. 
The expression in \cref{eq:MPS} shows that after choosing a basis, the MPS representation requires $\mathcal{O}(N\chi ^2d)$ complex numbers to be stored. 
Hence, provided that $\chi $ does not grow exponentially with system size, the MPS representation is efficient. Results from quantum information theory show that this is indeed the case~\cite{Vidal2003,Hastings2007,Eisert2010} for many relevant situations. 
\par Moreover, there exist efficient numerical algorithms for determining an MPS approximation for the ground state by variationally minimizing the energy expectation value $\mel*{\psi}{H}{\psi}$ for a given Hamiltonian $H$. 
To this end, the tensors $M_k^{i_k}$ are iteratively updated, one after another, while keeping all others fixed.
The optimal tensor in each step can be obtained by determining the ground state of an effective Hamiltonian describing the interaction of the site to be updated with its environment. 
Repeating the updating procedure, starting from the left boundary and sweeping back and forth until the relative change of the energy is below a certain tolerance $\varepsilon$, one obtains an MPS approximation $\ket*{\Omega}$ for the ground state~\cite{Schollwoeck2011}. 
After obtaining the ground state, the first excited states can be computed in a similar fashion by using the Hamiltonian $H+\alpha\ket*{\Omega}\bra{\Omega}$, where the constant $\alpha$ must be chosen to be large enough. 
In addition, there exist MPS methods for computing time evolution (up to certain time scales)~\cite{Schollwoeck2011,Orus2014a}. 
All of these algorithms do not suffer from the sign problem and have been successfully used to address many lattice models, in particular in regimes that are inaccessible with conventional MC methods. 
\par The MPS approach can be readily generalized to higher spatial dimensions~\cite{Verstraete2004b}, and there exist various other TN ans{\"a}tze for one and higher dimensions~\cite{Shi2006,Vidal2008,Evenbly2014,PhysRevLett.126.170603}.
Moreover, there exist first works using TNs in higher dimensions to explore lattice gauge theories~\cite{Felser2019,Magnifico2021}.
\begin{figure}[t!]
    \centering
    \includegraphics[width=0.99\columnwidth]{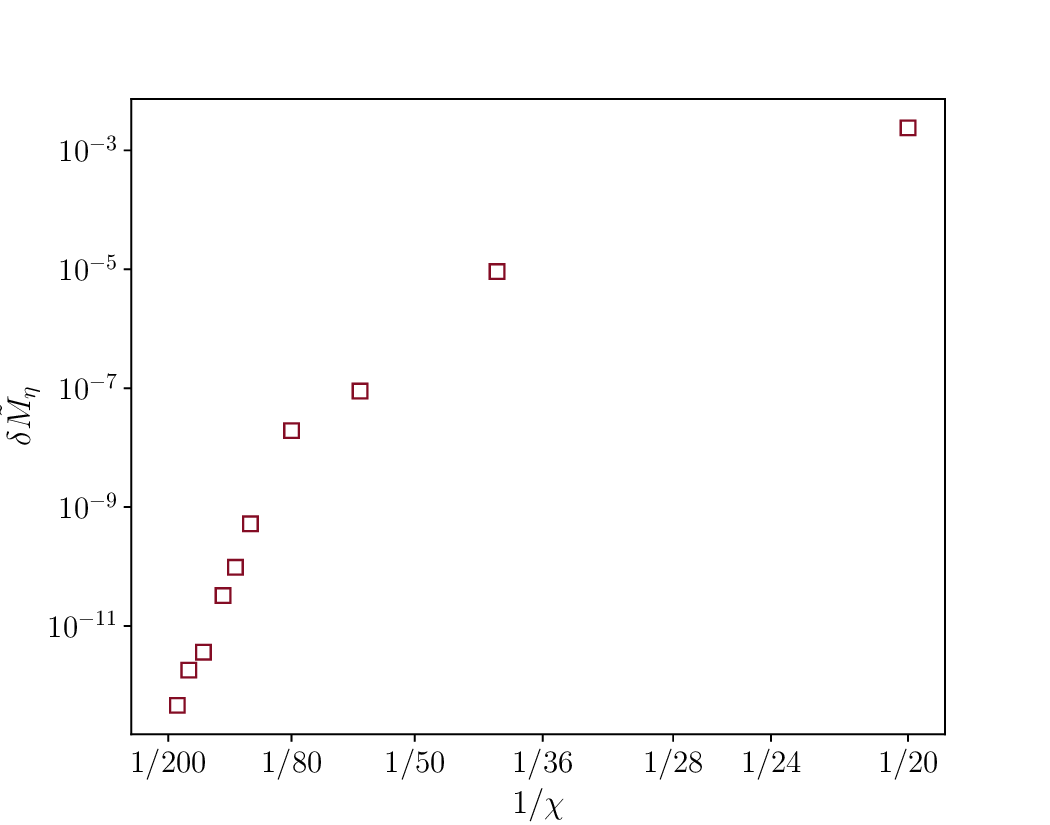}   
    \caption{An illustration of the behavior of the relative error in the meson mass as a function of the bond dimension $\chi$ for $(\tilde{a}\tilde{g},\tilde{m}/\tilde{g},\tilde{L}\tilde{g})=(0.5, 0.2, 16)$.
    \label{fig:extD}
    }
\end{figure}
In the MPS simulations of this work, $\varepsilon=10^{-8}$ and, for a given set of bare parameters $(\tilde{a}\tilde{g},\tilde{m}/\tilde{g},\tilde{L}\tilde{g})$, the variational computation of the ground state and the meson state are repeated for various values of $\chi$ starting from $20$ and increasing to a point where numerical errors due to finite $\chi$ become negligible. 
\cref{fig:extD} shows an example of the behavior of the relative error in the meson mass as a function of the bond dimension,  $\delta \tilde{M}_\eta(\chi) = \left[\tilde{M}_\eta(\chi)-\tilde{M}_\eta(200)\right]/\tilde{M}_\eta(200)$, where $\tilde{M}_\eta(\chi)$indicates the meson mass at a given $\chi$.
In general, similar behavior is observed for all the parameter sets $(\tilde{a}\tilde{g},\tilde{m}/\tilde{g},\tilde{L}\tilde{g})$ studied, and the relative error due to a finite bond dimension is insignificant for the precision targeted in this work.
\section{Comparison of meson mass scaling in Largrangian and Hamiltonian calculations\label{app:consistency-check}}
\par As a consistency check on the qualitative agreement between the Hamiltonian and Lagrangian systems used in this study, the scalings of the meson masses $M_\eta(m/g)$ and $\tilde{M}_\eta(\tilde{m}/\tilde{g})$ with the fermion mass are compared with each other in the vicinity of $m/g = 0.2$ and $\tilde{m}/\tilde{g}=0.2$, respectively.
Additionally, these scalings are compared with the corresponding analytic expression to first order in  continuum-limit mass perturbation theory~\cite{Adam1997},
    \begin{equation}
    \label{eq:continuum}
        \frac{M_\eta}{g} = \frac{1}{\sqrt{\pi}} \left(1 + 1.78105\sqrt{\pi} \left(\frac{m}{g}\right) + \mathcal{O}\!\left(\frac{m^2}{g^2}\right)\!\right)\thinspace.
    \end{equation}
\par To enable the comparison, Lagrangian gauge field configurations for several additional values of $m/g$ are generated. 
Specifically, in addition to the MC ensemble of $N_\mathrm{cfg} = 1000$ configurations at $(\tilde{a}\tilde{g}, \tilde{m}/\tilde{g}, \tilde{L}\tilde{g}) = (0.5, 0.2, 16)$, MC ensembles with $N_\mathrm{cfg} = 100$ configurations each are generated following an identical procedure, with $m/g \in \lbrace 0.02, 0.1, 0.4 \rbrace$, at the same values of $ag$ and $Lg$.
\par
\begin{figure}[t!]
    \centering
    \vspace{0.2pt}% otherwise figure slightly covers text
    \includegraphics[scale=0.54]{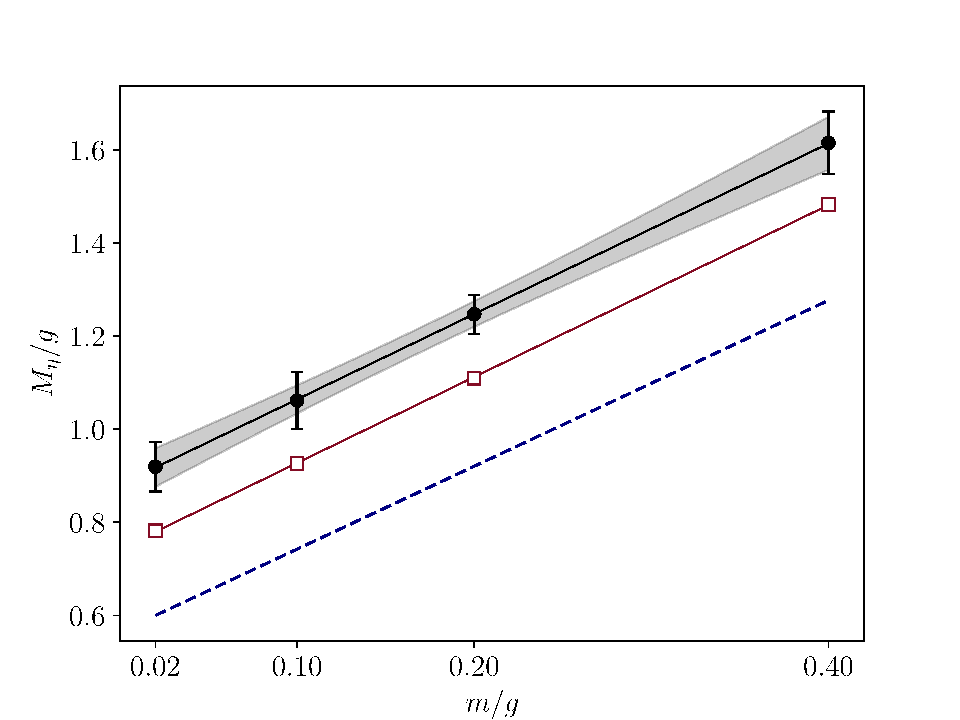}
    \caption{An illustration of the dependence of the meson mass on the bare fermion mass. 
    The continuum-limit analytic expression for $M_\eta$ in \cref{eq:continuum}, linearized in $m/g$, is indicated by a dashed blue line.
    The Lagrangian results obtained from MC calculations are denoted by black circles, and the Hamiltonian results from TN calculations are denoted by red empty squares.
    Linear fits to the results are shown as a black line with a gray band for the Lagrangian calculations, and a red line for the Hamiltonian calculations.
    Differences on the horizontal axis between the renormalized and bare values of $m$/$g$, and on both axes between the Hamiltonian and Lagrangian calculations, are neglected.
    }
    \label{fig:mass}
\end{figure}

 \cref{fig:mass} illustrates this comparison as realized between linear two-parameter fits of the MC-determined values of $ M_\eta(m/g)/g$ from all ensembles, those of the TN-determined values of $\tilde{M}_{\eta}(\tilde{m}/\tilde{g})/\tilde{g}$, and the linearized form of \cref{eq:continuum}.
The slope parameters of the two fits to MC and TN results are given by $1.8 \pm 0.2$ and $1.84 \pm 0.01$, with  $\chi^2/\mathrm{dof} = 5 \times 10^{-5}$ and $2 \times 10^{-6}$, respectively, indicating consistency between the Lagrangian and Hamiltonian formulations.
Furthermore, observables in both formulations are consistent with the continuum-limit value of $1.78105$ in \cref{eq:continuum} up to percent-level corrections which may be explained by discretization artifacts. 
Finally, the intercept parameters of the fits are consistent with the qualitative expectation that the Hamiltonian results for  $\tilde{M}_\eta/\tilde{g}$ with massless fermions are closer to the continuum-limit value of $1/\sqrt{\pi}$ than the corresponding Lagrangian results, since the Hamiltonian results already include the continuum limit in time.
%\bibliography{refs.bib} 

\begin{thebibliography}{93}%
\makeatletter
\providecommand \@ifxundefined [1]{%
 \@ifx{#1\undefined}
}%
\providecommand \@ifnum [1]{%
 \ifnum #1\expandafter \@firstoftwo
 \else \expandafter \@secondoftwo
 \fi
}%
\providecommand \@ifx [1]{%
 \ifx #1\expandafter \@firstoftwo
 \else \expandafter \@secondoftwo
 \fi
}%
\providecommand \natexlab [1]{#1}%
\providecommand \enquote  [1]{``#1''}%
\providecommand \bibnamefont  [1]{#1}%
\providecommand \bibfnamefont [1]{#1}%
\providecommand \citenamefont [1]{#1}%
\providecommand \href@noop [0]{\@secondoftwo}%
\providecommand \href [0]{\begingroup \@sanitize@url \@href}%
\providecommand \@href[1]{\@@startlink{#1}\@@href}%
\providecommand \@@href[1]{\endgroup#1\@@endlink}%
\providecommand \@sanitize@url [0]{\catcode `\\12\catcode `\$12\catcode
  `\&12\catcode `\#12\catcode `\^12\catcode `\_12\catcode `\%12\relax}%
\providecommand \@@startlink[1]{}%
\providecommand \@@endlink[0]{}%
\providecommand \url  [0]{\begingroup\@sanitize@url \@url }%
\providecommand \@url [1]{\endgroup\@href {#1}{\urlprefix }}%
\providecommand \urlprefix  [0]{URL }%
\providecommand \Eprint [0]{\href }%
\providecommand \doibase [0]{http://dx.doi.org/}%
\providecommand \selectlanguage [0]{\@gobble}%
\providecommand \bibinfo  [0]{\@secondoftwo}%
\providecommand \bibfield  [0]{\@secondoftwo}%
\providecommand \translation [1]{[#1]}%
\providecommand \BibitemOpen [0]{}%
\providecommand \bibitemStop [0]{}%
\providecommand \bibitemNoStop [0]{.\EOS\space}%
\providecommand \EOS [0]{\spacefactor3000\relax}%
\providecommand \BibitemShut  [1]{\csname bibitem#1\endcsname}%
\let\auto@bib@innerbib\@empty
%</preamble>
\bibitem [{\citenamefont {Wilson}(1974)}]{Wilson:1974sk}%
  \BibitemOpen
  \bibfield  {author} {\bibinfo {author} {\bibfnamefont {K.~G.}\ \bibnamefont
  {Wilson}},\ }\href {\doibase 10.1103/PhysRevD.10.2445} {\bibfield  {journal}
  {\bibinfo  {journal} {Phys. Rev. D}\ }\textbf {\bibinfo {volume} {10}},\
  \bibinfo {pages} {2445} (\bibinfo {year} {1974})}\BibitemShut {NoStop}%
\bibitem [{\citenamefont {Osterwalder}\ and\ \citenamefont
  {Seiler}(1978)}]{Osterwalder:1977pc}%
  \BibitemOpen
  \bibfield  {author} {\bibinfo {author} {\bibfnamefont {K.}~\bibnamefont
  {Osterwalder}}\ and\ \bibinfo {author} {\bibfnamefont {E.}~\bibnamefont
  {Seiler}},\ }\href {\doibase 10.1016/0003-4916(78)90039-8} {\bibfield
  {journal} {\bibinfo  {journal} {Annals Phys.}\ }\textbf {\bibinfo {volume}
  {110}},\ \bibinfo {pages} {440} (\bibinfo {year} {1978})}\BibitemShut
  {NoStop}%
\bibitem [{\citenamefont {Lehner}\ \emph {et~al.}(2019)\citenamefont {Lehner}
  \emph {et~al.}}]{USQCD:2019hyg}%
  \BibitemOpen
  \bibfield  {author} {\bibinfo {author} {\bibfnamefont {C.}~\bibnamefont
  {Lehner}} \emph {et~al.} (\bibinfo {collaboration} {USQCD}),\ }\href
  {\doibase 10.1140/epja/i2019-12891-2} {\bibfield  {journal} {\bibinfo
  {journal} {Eur. Phys. J. A}\ }\textbf {\bibinfo {volume} {55}},\ \bibinfo
  {pages} {195} (\bibinfo {year} {2019})},\ \Eprint
  {http://arxiv.org/abs/1904.09479} {arXiv:1904.09479 [hep-lat]} \BibitemShut
  {NoStop}%
\bibitem [{\citenamefont {Cirigliano}\ \emph {et~al.}(2019)\citenamefont
  {Cirigliano}, \citenamefont {Davoudi}, \citenamefont {Bhattacharya},
  \citenamefont {Izubuchi}, \citenamefont {Shanahan}, \citenamefont
  {Syritsyn},\ and\ \citenamefont {Wagman}}]{Cirigliano:2019jig}%
  \BibitemOpen
  \bibfield  {author} {\bibinfo {author} {\bibfnamefont {V.}~\bibnamefont
  {Cirigliano}}, \bibinfo {author} {\bibfnamefont {Z.}~\bibnamefont {Davoudi}},
  \bibinfo {author} {\bibfnamefont {T.}~\bibnamefont {Bhattacharya}}, \bibinfo
  {author} {\bibfnamefont {T.}~\bibnamefont {Izubuchi}}, \bibinfo {author}
  {\bibfnamefont {P.~E.}\ \bibnamefont {Shanahan}}, \bibinfo {author}
  {\bibfnamefont {S.}~\bibnamefont {Syritsyn}}, \ and\ \bibinfo {author}
  {\bibfnamefont {M.~L.}\ \bibnamefont {Wagman}} (\bibinfo {collaboration}
  {USQCD}),\ }\href {\doibase 10.1140/epja/i2019-12889-8} {\bibfield  {journal}
  {\bibinfo  {journal} {Eur. Phys. J. A}\ }\textbf {\bibinfo {volume} {55}},\
  \bibinfo {pages} {197} (\bibinfo {year} {2019})},\ \Eprint
  {http://arxiv.org/abs/1904.09704} {arXiv:1904.09704 [hep-lat]} \BibitemShut
  {NoStop}%
\bibitem [{\citenamefont {Brower}\ \emph {et~al.}(2019)\citenamefont {Brower},
  \citenamefont {Hasenfratz}, \citenamefont {Neil}, \citenamefont {Catterall},
  \citenamefont {Fleming}, \citenamefont {Giedt}, \citenamefont {Rinaldi},
  \citenamefont {Schaich}, \citenamefont {Weinberg},\ and\ \citenamefont
  {Witzel}}]{USQCD:2019hee}%
  \BibitemOpen
  \bibfield  {author} {\bibinfo {author} {\bibfnamefont {R.~C.}\ \bibnamefont
  {Brower}}, \bibinfo {author} {\bibfnamefont {A.}~\bibnamefont {Hasenfratz}},
  \bibinfo {author} {\bibfnamefont {E.~T.}\ \bibnamefont {Neil}}, \bibinfo
  {author} {\bibfnamefont {S.}~\bibnamefont {Catterall}}, \bibinfo {author}
  {\bibfnamefont {G.}~\bibnamefont {Fleming}}, \bibinfo {author} {\bibfnamefont
  {J.}~\bibnamefont {Giedt}}, \bibinfo {author} {\bibfnamefont
  {E.}~\bibnamefont {Rinaldi}}, \bibinfo {author} {\bibfnamefont
  {D.}~\bibnamefont {Schaich}}, \bibinfo {author} {\bibfnamefont
  {E.}~\bibnamefont {Weinberg}}, \ and\ \bibinfo {author} {\bibfnamefont
  {O.}~\bibnamefont {Witzel}} (\bibinfo {collaboration} {USQCD}),\ }\href
  {\doibase 10.1140/epja/i2019-12901-5} {\bibfield  {journal} {\bibinfo
  {journal} {Eur. Phys. J. A}\ }\textbf {\bibinfo {volume} {55}},\ \bibinfo
  {pages} {198} (\bibinfo {year} {2019})},\ \Eprint
  {http://arxiv.org/abs/1904.09964} {arXiv:1904.09964 [hep-lat]} \BibitemShut
  {NoStop}%
\bibitem [{\citenamefont {Aoki}\ \emph {et~al.}(2022)\citenamefont {Aoki} \emph
  {et~al.}}]{FlavourLatticeAveragingGroupFLAG:2021npn}%
  \BibitemOpen
  \bibfield  {author} {\bibinfo {author} {\bibfnamefont {Y.}~\bibnamefont
  {Aoki}} \emph {et~al.} (\bibinfo {collaboration} {Flavour Lattice Averaging
  Group (FLAG)}),\ }\href {\doibase 10.1140/epjc/s10052-022-10536-1} {\bibfield
   {journal} {\bibinfo  {journal} {Eur. Phys. J. C}\ }\textbf {\bibinfo
  {volume} {82}},\ \bibinfo {pages} {869} (\bibinfo {year} {2022})},\ \Eprint
  {http://arxiv.org/abs/2111.09849} {arXiv:2111.09849 [hep-lat]} \BibitemShut
  {NoStop}%
\bibitem [{\citenamefont {Kronfeld}\ \emph {et~al.}(2022)\citenamefont
  {Kronfeld} \emph {et~al.}}]{USQCD:2022mmc}%
  \BibitemOpen
  \bibfield  {author} {\bibinfo {author} {\bibfnamefont {A.~S.}\ \bibnamefont
  {Kronfeld}} \emph {et~al.} (\bibinfo {collaboration} {USQCD}),\ }\href@noop
  {} {\  (\bibinfo {year} {2022})},\ \Eprint {http://arxiv.org/abs/2207.07641}
  {arXiv:2207.07641 [hep-lat]} \BibitemShut {NoStop}%
\bibitem [{\citenamefont {Lin}\ \emph {et~al.}(2018)\citenamefont {Lin} \emph
  {et~al.}}]{Lin:2017snn}%
  \BibitemOpen
  \bibfield  {author} {\bibinfo {author} {\bibfnamefont {H.-W.}\ \bibnamefont
  {Lin}} \emph {et~al.},\ }\href {\doibase 10.1016/j.ppnp.2018.01.007}
  {\bibfield  {journal} {\bibinfo  {journal} {Prog. Part. Nucl. Phys.}\
  }\textbf {\bibinfo {volume} {100}},\ \bibinfo {pages} {107} (\bibinfo {year}
  {2018})},\ \Eprint {http://arxiv.org/abs/1711.07916} {arXiv:1711.07916
  [hep-ph]} \BibitemShut {NoStop}%
\bibitem [{\citenamefont {Bazavov}\ \emph {et~al.}(2019)\citenamefont
  {Bazavov}, \citenamefont {Karsch}, \citenamefont {Mukherjee},\ and\
  \citenamefont {Petreczky}}]{Bazavov:2019lgz}%
  \BibitemOpen
  \bibfield  {author} {\bibinfo {author} {\bibfnamefont {A.}~\bibnamefont
  {Bazavov}}, \bibinfo {author} {\bibfnamefont {F.}~\bibnamefont {Karsch}},
  \bibinfo {author} {\bibfnamefont {S.}~\bibnamefont {Mukherjee}}, \ and\
  \bibinfo {author} {\bibfnamefont {P.}~\bibnamefont {Petreczky}} (\bibinfo
  {collaboration} {USQCD}),\ }\href {\doibase 10.1140/epja/i2019-12922-0}
  {\bibfield  {journal} {\bibinfo  {journal} {Eur. Phys. J. A}\ }\textbf
  {\bibinfo {volume} {55}},\ \bibinfo {pages} {194} (\bibinfo {year} {2019})},\
  \Eprint {http://arxiv.org/abs/1904.09951} {arXiv:1904.09951 [hep-lat]}
  \BibitemShut {NoStop}%
\bibitem [{\citenamefont {Kronfeld}\ \emph {et~al.}(2019)\citenamefont
  {Kronfeld}, \citenamefont {Richards}, \citenamefont {Detmold}, \citenamefont
  {Gupta}, \citenamefont {Lin}, \citenamefont {Liu}, \citenamefont {Meyer},
  \citenamefont {Sufian},\ and\ \citenamefont {Syritsyn}}]{Kronfeld:2019nfb}%
  \BibitemOpen
  \bibfield  {author} {\bibinfo {author} {\bibfnamefont {A.~S.}\ \bibnamefont
  {Kronfeld}}, \bibinfo {author} {\bibfnamefont {D.~G.}\ \bibnamefont
  {Richards}}, \bibinfo {author} {\bibfnamefont {W.}~\bibnamefont {Detmold}},
  \bibinfo {author} {\bibfnamefont {R.}~\bibnamefont {Gupta}}, \bibinfo
  {author} {\bibfnamefont {H.-W.}\ \bibnamefont {Lin}}, \bibinfo {author}
  {\bibfnamefont {K.-F.}\ \bibnamefont {Liu}}, \bibinfo {author} {\bibfnamefont
  {A.~S.}\ \bibnamefont {Meyer}}, \bibinfo {author} {\bibfnamefont
  {R.}~\bibnamefont {Sufian}}, \ and\ \bibinfo {author} {\bibfnamefont
  {S.}~\bibnamefont {Syritsyn}} (\bibinfo {collaboration} {USQCD}),\ }\href
  {\doibase 10.1140/epja/i2019-12916-x} {\bibfield  {journal} {\bibinfo
  {journal} {Eur. Phys. J. A}\ }\textbf {\bibinfo {volume} {55}},\ \bibinfo
  {pages} {196} (\bibinfo {year} {2019})},\ \Eprint
  {http://arxiv.org/abs/1904.09931} {arXiv:1904.09931 [hep-lat]} \BibitemShut
  {NoStop}%
\bibitem [{\citenamefont {Amoroso}\ \emph {et~al.}(2022)\citenamefont {Amoroso}
  \emph {et~al.}}]{Amoroso:2022eow}%
  \BibitemOpen
  \bibfield  {author} {\bibinfo {author} {\bibfnamefont {S.}~\bibnamefont
  {Amoroso}} \emph {et~al.},\ }\href@noop {} {\  (\bibinfo {year} {2022})},\
  \Eprint {http://arxiv.org/abs/2203.13923} {arXiv:2203.13923 [hep-ph]}
  \BibitemShut {NoStop}%
\bibitem [{\citenamefont {Parisi}(1984)}]{Parisi:1983ae}%
  \BibitemOpen
  \bibfield  {author} {\bibinfo {author} {\bibfnamefont {G.}~\bibnamefont
  {Parisi}},\ }\href {\doibase 10.1016/0370-1573(84)90081-4} {\bibfield
  {journal} {\bibinfo  {journal} {Phys. Rept.}\ }\textbf {\bibinfo {volume}
  {103}},\ \bibinfo {pages} {203} (\bibinfo {year} {1984})}\BibitemShut
  {NoStop}%
\bibitem [{\citenamefont {Gibbs}(1986)}]{Gibbs:1986ut}%
  \BibitemOpen
  \bibfield  {author} {\bibinfo {author} {\bibfnamefont {P.~E.}\ \bibnamefont
  {Gibbs}},\ }\href {\doibase 10.1016/0370-2693(86)90109-7} {\bibfield
  {journal} {\bibinfo  {journal} {Phys. Lett. B}\ }\textbf {\bibinfo {volume}
  {182}},\ \bibinfo {pages} {369} (\bibinfo {year} {1986})}\BibitemShut
  {NoStop}%
\bibitem [{\citenamefont {Lepage}(1989)}]{Lepage:1989hd}%
  \BibitemOpen
  \bibfield  {author} {\bibinfo {author} {\bibfnamefont {G.~P.}\ \bibnamefont
  {Lepage}},\ }in\ \href@noop {} {\emph {\bibinfo {booktitle} {{Theoretical
  Advanced Study Institute in Elementary Particle Physics}}}}\ (\bibinfo {year}
  {1989})\BibitemShut {NoStop}%
\bibitem [{\citenamefont {Ba\~nuls}\ \emph {et~al.}(2018)\citenamefont
  {Ba\~nuls}, \citenamefont {Cichy}, \citenamefont {Cirac}, \citenamefont
  {Jansen},\ and\ \citenamefont {K\"uhn}}]{Banuls:2018jag}%
  \BibitemOpen
  \bibfield  {author} {\bibinfo {author} {\bibfnamefont {M.~C.}\ \bibnamefont
  {Ba\~nuls}}, \bibinfo {author} {\bibfnamefont {K.}~\bibnamefont {Cichy}},
  \bibinfo {author} {\bibfnamefont {J.~I.}\ \bibnamefont {Cirac}}, \bibinfo
  {author} {\bibfnamefont {K.}~\bibnamefont {Jansen}}, \ and\ \bibinfo {author}
  {\bibfnamefont {S.}~\bibnamefont {K\"uhn}},\ }\href {\doibase
  10.22323/1.334.0022} {\bibfield  {journal} {\bibinfo  {journal} {PoS}\
  }\textbf {\bibinfo {volume} {LATTICE2018}},\ \bibinfo {pages} {022} (\bibinfo
  {year} {2018})},\ \Eprint {http://arxiv.org/abs/1810.12838} {arXiv:1810.12838
  [hep-lat]} \BibitemShut {NoStop}%
\bibitem [{\citenamefont {Ba\~nuls}\ and\ \citenamefont
  {Cichy}(2020)}]{Banuls:2019rao}%
  \BibitemOpen
  \bibfield  {author} {\bibinfo {author} {\bibfnamefont {M.~C.}\ \bibnamefont
  {Ba\~nuls}}\ and\ \bibinfo {author} {\bibfnamefont {K.}~\bibnamefont
  {Cichy}},\ }\href {\doibase 10.1088/1361-6633/ab6311} {\bibfield  {journal}
  {\bibinfo  {journal} {Rept. Prog. Phys.}\ }\textbf {\bibinfo {volume} {83}},\
  \bibinfo {pages} {024401} (\bibinfo {year} {2020})},\ \Eprint
  {http://arxiv.org/abs/1910.00257} {arXiv:1910.00257 [hep-lat]} \BibitemShut
  {NoStop}%
\bibitem [{\citenamefont {Ba\~nuls}\ \emph {et~al.}(2020)\citenamefont
  {Ba\~nuls} \emph {et~al.}}]{Banuls:2019bmf}%
  \BibitemOpen
  \bibfield  {author} {\bibinfo {author} {\bibfnamefont {M.~C.}\ \bibnamefont
  {Ba\~nuls}} \emph {et~al.},\ }\href {\doibase 10.1140/epjd/e2020-100571-8}
  {\bibfield  {journal} {\bibinfo  {journal} {Eur. Phys. J. D}\ }\textbf
  {\bibinfo {volume} {74}},\ \bibinfo {pages} {165} (\bibinfo {year} {2020})},\
  \Eprint {http://arxiv.org/abs/1911.00003} {arXiv:1911.00003 [quant-ph]}
  \BibitemShut {NoStop}%
\bibitem [{\citenamefont {Haase}\ \emph {et~al.}(2021)\citenamefont {Haase},
  \citenamefont {Dellantonio}, \citenamefont {Celi}, \citenamefont {Paulson},
  \citenamefont {Kan}, \citenamefont {Jansen},\ and\ \citenamefont
  {Muschik}}]{Haase:2020kaj}%
  \BibitemOpen
  \bibfield  {author} {\bibinfo {author} {\bibfnamefont {J.~F.}\ \bibnamefont
  {Haase}}, \bibinfo {author} {\bibfnamefont {L.}~\bibnamefont {Dellantonio}},
  \bibinfo {author} {\bibfnamefont {A.}~\bibnamefont {Celi}}, \bibinfo {author}
  {\bibfnamefont {D.}~\bibnamefont {Paulson}}, \bibinfo {author} {\bibfnamefont
  {A.}~\bibnamefont {Kan}}, \bibinfo {author} {\bibfnamefont {K.}~\bibnamefont
  {Jansen}}, \ and\ \bibinfo {author} {\bibfnamefont {C.~A.}\ \bibnamefont
  {Muschik}},\ }\href {\doibase 10.22331/q-2021-02-04-393} {\bibfield
  {journal} {\bibinfo  {journal} {Quantum}\ }\textbf {\bibinfo {volume} {5}},\
  \bibinfo {pages} {393} (\bibinfo {year} {2021})},\ \Eprint
  {http://arxiv.org/abs/2006.14160} {arXiv:2006.14160 [quant-ph]} \BibitemShut
  {NoStop}%
\bibitem [{\citenamefont {Bauer}\ \emph
  {et~al.}(2023{\natexlab{a}})\citenamefont {Bauer} \emph
  {et~al.}}]{Bauer:2022hpo}%
  \BibitemOpen
  \bibfield  {author} {\bibinfo {author} {\bibfnamefont {C.~W.}\ \bibnamefont
  {Bauer}} \emph {et~al.},\ }\href {\doibase 10.1103/PRXQuantum.4.027001}
  {\bibfield  {journal} {\bibinfo  {journal} {PRX Quantum}\ }\textbf {\bibinfo
  {volume} {4}},\ \bibinfo {pages} {027001} (\bibinfo {year}
  {2023}{\natexlab{a}})},\ \Eprint {http://arxiv.org/abs/2204.03381}
  {arXiv:2204.03381 [quant-ph]} \BibitemShut {NoStop}%
\bibitem [{\citenamefont {Meurice}\ \emph {et~al.}(2022)\citenamefont
  {Meurice}, \citenamefont {Osborn}, \citenamefont {Sakai}, \citenamefont
  {Unmuth-Yockey}, \citenamefont {Catterall},\ and\ \citenamefont
  {Somma}}]{Meurice:2022xbk}%
  \BibitemOpen
  \bibfield  {author} {\bibinfo {author} {\bibfnamefont {Y.}~\bibnamefont
  {Meurice}}, \bibinfo {author} {\bibfnamefont {J.~C.}\ \bibnamefont {Osborn}},
  \bibinfo {author} {\bibfnamefont {R.}~\bibnamefont {Sakai}}, \bibinfo
  {author} {\bibfnamefont {J.}~\bibnamefont {Unmuth-Yockey}}, \bibinfo {author}
  {\bibfnamefont {S.}~\bibnamefont {Catterall}}, \ and\ \bibinfo {author}
  {\bibfnamefont {R.~D.}\ \bibnamefont {Somma}},\ }in\ \href@noop {} {\emph
  {\bibinfo {booktitle} {{Snowmass 2021}}}}\ (\bibinfo {year} {2022})\ \Eprint
  {http://arxiv.org/abs/2203.04902} {arXiv:2203.04902 [hep-lat]} \BibitemShut
  {NoStop}%
\bibitem [{\citenamefont {Catterall}\ \emph {et~al.}(2022)\citenamefont
  {Catterall} \emph {et~al.}}]{Catterall:2022wjq}%
  \BibitemOpen
  \bibfield  {author} {\bibinfo {author} {\bibfnamefont {S.}~\bibnamefont
  {Catterall}} \emph {et~al.},\ }in\ \href@noop {} {\emph {\bibinfo {booktitle}
  {{Snowmass 2021}}}}\ (\bibinfo {year} {2022})\ \Eprint
  {http://arxiv.org/abs/2209.14839} {arXiv:2209.14839 [quant-ph]} \BibitemShut
  {NoStop}%
\bibitem [{\citenamefont {Di~Meglio}\ \emph {et~al.}(2023)\citenamefont
  {Di~Meglio} \emph {et~al.}}]{DiMeglio:2023nsa}%
  \BibitemOpen
  \bibfield  {author} {\bibinfo {author} {\bibfnamefont {A.}~\bibnamefont
  {Di~Meglio}} \emph {et~al.},\ }\href@noop {} {\  (\bibinfo {year} {2023})},\
  \Eprint {http://arxiv.org/abs/2307.03236} {arXiv:2307.03236 [quant-ph]}
  \BibitemShut {NoStop}%
\bibitem [{\citenamefont {Bauer}\ \emph
  {et~al.}(2023{\natexlab{b}})\citenamefont {Bauer}, \citenamefont {Davoudi},
  \citenamefont {Klco},\ and\ \citenamefont {Savage}}]{Bauer:2023qgm}%
  \BibitemOpen
  \bibfield  {author} {\bibinfo {author} {\bibfnamefont {C.~W.}\ \bibnamefont
  {Bauer}}, \bibinfo {author} {\bibfnamefont {Z.}~\bibnamefont {Davoudi}},
  \bibinfo {author} {\bibfnamefont {N.}~\bibnamefont {Klco}}, \ and\ \bibinfo
  {author} {\bibfnamefont {M.~J.}\ \bibnamefont {Savage}},\ }\href {\doibase
  10.1038/s42254-023-00599-8} {\bibfield  {journal} {\bibinfo  {journal}
  {Nature Rev. Phys.}\ }\textbf {\bibinfo {volume} {5}},\ \bibinfo {pages}
  {420} (\bibinfo {year} {2023}{\natexlab{b}})},\ \Eprint
  {http://arxiv.org/abs/2404.06298} {arXiv:2404.06298 [hep-ph]} \BibitemShut
  {NoStop}%
\bibitem [{\citenamefont {Ba{\~n}uls}\ \emph {et~al.}(2019)\citenamefont
  {Ba{\~n}uls}, \citenamefont {Cichy}, \citenamefont {Cirac}, \citenamefont
  {Jansen},\ and\ \citenamefont {Kühn}}]{Banuls2018a}%
  \BibitemOpen
  \bibfield  {author} {\bibinfo {author} {\bibfnamefont {M.~C.}\ \bibnamefont
  {Ba{\~n}uls}}, \bibinfo {author} {\bibfnamefont {K.}~\bibnamefont {Cichy}},
  \bibinfo {author} {\bibfnamefont {J.~I.}\ \bibnamefont {Cirac}}, \bibinfo
  {author} {\bibfnamefont {K.}~\bibnamefont {Jansen}}, \ and\ \bibinfo {author}
  {\bibfnamefont {S.}~\bibnamefont {Kühn}},\ }\href {\doibase
  10.22323/1.334.0022} {\bibfield  {journal} {\bibinfo  {journal} {PoS(LATTICE
  2018)022}\ } (\bibinfo {year} {2019}),\ 10.22323/1.334.0022}\BibitemShut
  {NoStop}%
\bibitem [{\citenamefont {Ba{\~{n}}uls}\ and\ \citenamefont
  {Cichy}(2020)}]{Banuls2019}%
  \BibitemOpen
  \bibfield  {author} {\bibinfo {author} {\bibfnamefont {M.~C.}\ \bibnamefont
  {Ba{\~{n}}uls}}\ and\ \bibinfo {author} {\bibfnamefont {K.}~\bibnamefont
  {Cichy}},\ }\href {\doibase 10.1088/1361-6633/ab6311} {\bibfield  {journal}
  {\bibinfo  {journal} {Rep. Prog. Phys.}\ }\textbf {\bibinfo {volume} {83}},\
  \bibinfo {pages} {024401} (\bibinfo {year} {2020})}\BibitemShut {NoStop}%
\bibitem [{\citenamefont {Ba{\~{n}}uls}\ \emph {et~al.}(2020)\citenamefont
  {Ba{\~{n}}uls}, \citenamefont {Blatt}, \citenamefont {Catani}, \citenamefont
  {Celi}, \citenamefont {Cirac}, \citenamefont {Dalmonte}, \citenamefont
  {Fallani}, \citenamefont {Jansen}, \citenamefont {Lewenstein}, \citenamefont
  {Montangero}, \citenamefont {Muschik}, \citenamefont {Reznik}, \citenamefont
  {Rico}, \citenamefont {Tagliacozzo}, \citenamefont {Acoleyen}, \citenamefont
  {Verstraete}, \citenamefont {Wiese}, \citenamefont {Wingate}, \citenamefont
  {Zakrzewski},\ and\ \citenamefont {Zoller}}]{Banuls2020}%
  \BibitemOpen
  \bibfield  {author} {\bibinfo {author} {\bibfnamefont {M.~C.}\ \bibnamefont
  {Ba{\~{n}}uls}}, \bibinfo {author} {\bibfnamefont {R.}~\bibnamefont {Blatt}},
  \bibinfo {author} {\bibfnamefont {J.}~\bibnamefont {Catani}}, \bibinfo
  {author} {\bibfnamefont {A.}~\bibnamefont {Celi}}, \bibinfo {author}
  {\bibfnamefont {J.~I.}\ \bibnamefont {Cirac}}, \bibinfo {author}
  {\bibfnamefont {M.}~\bibnamefont {Dalmonte}}, \bibinfo {author}
  {\bibfnamefont {L.}~\bibnamefont {Fallani}}, \bibinfo {author} {\bibfnamefont
  {K.}~\bibnamefont {Jansen}}, \bibinfo {author} {\bibfnamefont
  {M.}~\bibnamefont {Lewenstein}}, \bibinfo {author} {\bibfnamefont
  {S.}~\bibnamefont {Montangero}}, \bibinfo {author} {\bibfnamefont {C.~A.}\
  \bibnamefont {Muschik}}, \bibinfo {author} {\bibfnamefont {B.}~\bibnamefont
  {Reznik}}, \bibinfo {author} {\bibfnamefont {E.}~\bibnamefont {Rico}},
  \bibinfo {author} {\bibfnamefont {L.}~\bibnamefont {Tagliacozzo}}, \bibinfo
  {author} {\bibfnamefont {K.~V.}\ \bibnamefont {Acoleyen}}, \bibinfo {author}
  {\bibfnamefont {F.}~\bibnamefont {Verstraete}}, \bibinfo {author}
  {\bibfnamefont {U.-J.}\ \bibnamefont {Wiese}}, \bibinfo {author}
  {\bibfnamefont {M.}~\bibnamefont {Wingate}}, \bibinfo {author} {\bibfnamefont
  {J.}~\bibnamefont {Zakrzewski}}, \ and\ \bibinfo {author} {\bibfnamefont
  {P.}~\bibnamefont {Zoller}},\ }\href {\doibase 10.1140/epjd/e2020-100571-8}
  {\bibfield  {journal} {\bibinfo  {journal} {The European Physical Journal D}\
  }\textbf {\bibinfo {volume} {74}} (\bibinfo {year} {2020}),\
  10.1140/epjd/e2020-100571-8}\BibitemShut {NoStop}%
\bibitem [{\citenamefont {Funcke}\ \emph
  {et~al.}(2023{\natexlab{a}})\citenamefont {Funcke}, \citenamefont {Hartung},
  \citenamefont {Jansen},\ and\ \citenamefont {Kühn}}]{Funcke2023a}%
  \BibitemOpen
  \bibfield  {author} {\bibinfo {author} {\bibfnamefont {L.}~\bibnamefont
  {Funcke}}, \bibinfo {author} {\bibfnamefont {T.}~\bibnamefont {Hartung}},
  \bibinfo {author} {\bibfnamefont {K.}~\bibnamefont {Jansen}}, \ and\ \bibinfo
  {author} {\bibfnamefont {S.}~\bibnamefont {Kühn}},\ }\href {\doibase
  10.22323/1.430.0228} {\bibfield  {journal} {\bibinfo  {journal} {PoS}\
  }\textbf {\bibinfo {volume} {(LATTICE2022)}},\ \bibinfo {pages} {228}
  (\bibinfo {year} {2023}{\natexlab{a}})}\BibitemShut {NoStop}%
\bibitem [{\citenamefont {Magnifico}\ \emph {et~al.}(2021)\citenamefont
  {Magnifico}, \citenamefont {Felser}, \citenamefont {Silvi},\ and\
  \citenamefont {Montangero}}]{Magnifico2021}%
  \BibitemOpen
  \bibfield  {author} {\bibinfo {author} {\bibfnamefont {G.}~\bibnamefont
  {Magnifico}}, \bibinfo {author} {\bibfnamefont {T.}~\bibnamefont {Felser}},
  \bibinfo {author} {\bibfnamefont {P.}~\bibnamefont {Silvi}}, \ and\ \bibinfo
  {author} {\bibfnamefont {S.}~\bibnamefont {Montangero}},\ }\href {\doibase
  10.1038/s41467-021-23646-3} {\bibfield  {journal} {\bibinfo  {journal} {Nat.
  Commun.}\ }\textbf {\bibinfo {volume} {12}},\ \bibinfo {pages} {3600}
  (\bibinfo {year} {2021})}\BibitemShut {NoStop}%
\bibitem [{\citenamefont {Felser}\ \emph {et~al.}(2020)\citenamefont {Felser},
  \citenamefont {Silvi}, \citenamefont {Collura},\ and\ \citenamefont
  {Montangero}}]{Felser2019}%
  \BibitemOpen
  \bibfield  {author} {\bibinfo {author} {\bibfnamefont {T.}~\bibnamefont
  {Felser}}, \bibinfo {author} {\bibfnamefont {P.}~\bibnamefont {Silvi}},
  \bibinfo {author} {\bibfnamefont {M.}~\bibnamefont {Collura}}, \ and\
  \bibinfo {author} {\bibfnamefont {S.}~\bibnamefont {Montangero}},\ }\href
  {\doibase 10.1103/PhysRevX.10.041040} {\bibfield  {journal} {\bibinfo
  {journal} {Phys. Rev. X}\ }\textbf {\bibinfo {volume} {10}},\ \bibinfo
  {pages} {041040} (\bibinfo {year} {2020})}\BibitemShut {NoStop}%
\bibitem [{\citenamefont {Avkhadiev}\ \emph {et~al.}(2020)\citenamefont
  {Avkhadiev}, \citenamefont {Shanahan},\ and\ \citenamefont
  {Young}}]{Avkhadiev:2019niu}%
  \BibitemOpen
  \bibfield  {author} {\bibinfo {author} {\bibfnamefont {A.}~\bibnamefont
  {Avkhadiev}}, \bibinfo {author} {\bibfnamefont {P.~E.}\ \bibnamefont
  {Shanahan}}, \ and\ \bibinfo {author} {\bibfnamefont {R.~D.}\ \bibnamefont
  {Young}},\ }\href {\doibase 10.1103/PhysRevLett.124.080501} {\bibfield
  {journal} {\bibinfo  {journal} {Phys. Rev. Lett.}\ }\textbf {\bibinfo
  {volume} {124}},\ \bibinfo {pages} {080501} (\bibinfo {year} {2020})},\
  \Eprint {http://arxiv.org/abs/1908.04194} {arXiv:1908.04194 [hep-lat]}
  \BibitemShut {NoStop}%
\bibitem [{\citenamefont {Harmalkar}\ \emph {et~al.}(2020)\citenamefont
  {Harmalkar}, \citenamefont {Lamm},\ and\ \citenamefont
  {Lawrence}}]{Harmalkar:2020mpd}%
  \BibitemOpen
  \bibfield  {author} {\bibinfo {author} {\bibfnamefont {S.}~\bibnamefont
  {Harmalkar}}, \bibinfo {author} {\bibfnamefont {H.}~\bibnamefont {Lamm}}, \
  and\ \bibinfo {author} {\bibfnamefont {S.}~\bibnamefont {Lawrence}} (\bibinfo
  {collaboration} {NuQS}),\ }\href@noop {} {\  (\bibinfo {year} {2020})},\
  \Eprint {http://arxiv.org/abs/2001.11490} {arXiv:2001.11490 [hep-lat]}
  \BibitemShut {NoStop}%
\bibitem [{\citenamefont {Avkhadiev}\ \emph {et~al.}(2023)\citenamefont
  {Avkhadiev}, \citenamefont {Shanahan},\ and\ \citenamefont
  {Young}}]{Avkhadiev:2022ttx}%
  \BibitemOpen
  \bibfield  {author} {\bibinfo {author} {\bibfnamefont {A.}~\bibnamefont
  {Avkhadiev}}, \bibinfo {author} {\bibfnamefont {P.~E.}\ \bibnamefont
  {Shanahan}}, \ and\ \bibinfo {author} {\bibfnamefont {R.~D.}\ \bibnamefont
  {Young}},\ }\href {\doibase 10.1103/PhysRevD.107.054507} {\bibfield
  {journal} {\bibinfo  {journal} {Phys. Rev. D}\ }\textbf {\bibinfo {volume}
  {107}},\ \bibinfo {pages} {054507} (\bibinfo {year} {2023})},\ \Eprint
  {http://arxiv.org/abs/2209.01209} {arXiv:2209.01209 [hep-lat]} \BibitemShut
  {NoStop}%
\bibitem [{\citenamefont {Clemente}\ \emph {et~al.}(2022)\citenamefont
  {Clemente}, \citenamefont {Crippa},\ and\ \citenamefont
  {Jansen}}]{Clemente:2022cka}%
  \BibitemOpen
  \bibfield  {author} {\bibinfo {author} {\bibfnamefont {G.}~\bibnamefont
  {Clemente}}, \bibinfo {author} {\bibfnamefont {A.}~\bibnamefont {Crippa}}, \
  and\ \bibinfo {author} {\bibfnamefont {K.}~\bibnamefont {Jansen}},\ }\href
  {\doibase 10.1103/PhysRevD.106.114511} {\bibfield  {journal} {\bibinfo
  {journal} {Phys. Rev. D}\ }\textbf {\bibinfo {volume} {106}},\ \bibinfo
  {pages} {114511} (\bibinfo {year} {2022})}\BibitemShut {NoStop}%
\bibitem [{\citenamefont {Carena}\ \emph {et~al.}(2022)\citenamefont {Carena},
  \citenamefont {Gustafson}, \citenamefont {Lamm}, \citenamefont {Li},\ and\
  \citenamefont {Liu}}]{Carena:2022hpz}%
  \BibitemOpen
  \bibfield  {author} {\bibinfo {author} {\bibfnamefont {M.}~\bibnamefont
  {Carena}}, \bibinfo {author} {\bibfnamefont {E.~J.}\ \bibnamefont
  {Gustafson}}, \bibinfo {author} {\bibfnamefont {H.}~\bibnamefont {Lamm}},
  \bibinfo {author} {\bibfnamefont {Y.-Y.}\ \bibnamefont {Li}}, \ and\ \bibinfo
  {author} {\bibfnamefont {W.}~\bibnamefont {Liu}},\ }\href@noop {} {\
  (\bibinfo {year} {2022})},\ \Eprint {http://arxiv.org/abs/2208.10417}
  {arXiv:2208.10417 [hep-lat]} \BibitemShut {NoStop}%
\bibitem [{\citenamefont {Funcke}\ \emph
  {et~al.}(2023{\natexlab{b}})\citenamefont {Funcke}, \citenamefont {Hartung},
  \citenamefont {Jansen},\ and\ \citenamefont {K\"uhn}}]{Funcke:2023jbq}%
  \BibitemOpen
  \bibfield  {author} {\bibinfo {author} {\bibfnamefont {L.}~\bibnamefont
  {Funcke}}, \bibinfo {author} {\bibfnamefont {T.}~\bibnamefont {Hartung}},
  \bibinfo {author} {\bibfnamefont {K.}~\bibnamefont {Jansen}}, \ and\ \bibinfo
  {author} {\bibfnamefont {S.}~\bibnamefont {K\"uhn}},\ }\href {\doibase
  10.22323/1.430.0228} {\bibfield  {journal} {\bibinfo  {journal} {PoS}\
  }\textbf {\bibinfo {volume} {LATTICE2022}},\ \bibinfo {pages} {228} (\bibinfo
  {year} {2023}{\natexlab{b}})},\ \Eprint {http://arxiv.org/abs/2302.00467}
  {arXiv:2302.00467 [hep-lat]} \BibitemShut {NoStop}%
\bibitem [{\citenamefont {Crippa}\ \emph {et~al.}(2024)\citenamefont {Crippa},
  \citenamefont {Romiti}, \citenamefont {Funcke}, \citenamefont {Jansen},
  \citenamefont {K\"uhn}, \citenamefont {Stornati},\ and\ \citenamefont
  {Urbach}}]{Crippa:2024cqr}%
  \BibitemOpen
  \bibfield  {author} {\bibinfo {author} {\bibfnamefont {A.}~\bibnamefont
  {Crippa}}, \bibinfo {author} {\bibfnamefont {S.}~\bibnamefont {Romiti}},
  \bibinfo {author} {\bibfnamefont {L.}~\bibnamefont {Funcke}}, \bibinfo
  {author} {\bibfnamefont {K.}~\bibnamefont {Jansen}}, \bibinfo {author}
  {\bibfnamefont {S.}~\bibnamefont {K\"uhn}}, \bibinfo {author} {\bibfnamefont
  {P.}~\bibnamefont {Stornati}}, \ and\ \bibinfo {author} {\bibfnamefont
  {C.}~\bibnamefont {Urbach}},\ }\href@noop {} {\  (\bibinfo {year} {2024})},\
  \Eprint {http://arxiv.org/abs/2404.17545} {arXiv:2404.17545 [hep-lat]}
  \BibitemShut {NoStop}%
\bibitem [{\citenamefont {Wagman}(2024)}]{Wagman:2024rid}%
  \BibitemOpen
  \bibfield  {author} {\bibinfo {author} {\bibfnamefont {M.~L.}\ \bibnamefont
  {Wagman}},\ }\href@noop {} {\  (\bibinfo {year} {2024})},\ \Eprint
  {http://arxiv.org/abs/2406.20009} {arXiv:2406.20009 [hep-lat]} \BibitemShut
  {NoStop}%
\bibitem [{\citenamefont {Hackett}\ and\ \citenamefont
  {Wagman}(2024)}]{Hackett:2024xnx}%
  \BibitemOpen
  \bibfield  {author} {\bibinfo {author} {\bibfnamefont {D.~C.}\ \bibnamefont
  {Hackett}}\ and\ \bibinfo {author} {\bibfnamefont {M.~L.}\ \bibnamefont
  {Wagman}},\ }\href@noop {} {\  (\bibinfo {year} {2024})},\ \Eprint
  {http://arxiv.org/abs/2407.21777} {arXiv:2407.21777 [hep-lat]} \BibitemShut
  {NoStop}%
\bibitem [{\citenamefont {Albanese}\ \emph {et~al.}(1987)\citenamefont
  {Albanese} \emph {et~al.}}]{APE:1987ehd}%
  \BibitemOpen
  \bibfield  {author} {\bibinfo {author} {\bibfnamefont {M.}~\bibnamefont
  {Albanese}} \emph {et~al.} (\bibinfo {collaboration} {APE}),\ }\href
  {\doibase 10.1016/0370-2693(87)91160-9} {\bibfield  {journal} {\bibinfo
  {journal} {Phys. Lett. B}\ }\textbf {\bibinfo {volume} {192}},\ \bibinfo
  {pages} {163} (\bibinfo {year} {1987})}\BibitemShut {NoStop}%
\bibitem [{\citenamefont {Falcioni}\ \emph {et~al.}(1985)\citenamefont
  {Falcioni}, \citenamefont {Paciello}, \citenamefont {Parisi},\ and\
  \citenamefont {Taglienti}}]{Falcioni:1984ei}%
  \BibitemOpen
  \bibfield  {author} {\bibinfo {author} {\bibfnamefont {M.}~\bibnamefont
  {Falcioni}}, \bibinfo {author} {\bibfnamefont {M.~L.}\ \bibnamefont
  {Paciello}}, \bibinfo {author} {\bibfnamefont {G.}~\bibnamefont {Parisi}}, \
  and\ \bibinfo {author} {\bibfnamefont {B.}~\bibnamefont {Taglienti}},\ }\href
  {\doibase 10.1016/0550-3213(85)90280-9} {\bibfield  {journal} {\bibinfo
  {journal} {Nucl. Phys. B}\ }\textbf {\bibinfo {volume} {251}},\ \bibinfo
  {pages} {624} (\bibinfo {year} {1985})}\BibitemShut {NoStop}%
\bibitem [{\citenamefont {Teper}(1987)}]{Teper:1987wt}%
  \BibitemOpen
  \bibfield  {author} {\bibinfo {author} {\bibfnamefont {M.}~\bibnamefont
  {Teper}},\ }\href {\doibase 10.1016/0370-2693(87)90976-2} {\bibfield
  {journal} {\bibinfo  {journal} {Phys. Lett. B}\ }\textbf {\bibinfo {volume}
  {183}},\ \bibinfo {pages} {345} (\bibinfo {year} {1987})}\BibitemShut
  {NoStop}%
\bibitem [{\citenamefont {Morningstar}\ and\ \citenamefont
  {Peardon}(2004)}]{Morningstar:2003gk}%
  \BibitemOpen
  \bibfield  {author} {\bibinfo {author} {\bibfnamefont {C.}~\bibnamefont
  {Morningstar}}\ and\ \bibinfo {author} {\bibfnamefont {M.~J.}\ \bibnamefont
  {Peardon}},\ }\href {\doibase 10.1103/PhysRevD.69.054501} {\bibfield
  {journal} {\bibinfo  {journal} {Phys. Rev. D}\ }\textbf {\bibinfo {volume}
  {69}},\ \bibinfo {pages} {054501} (\bibinfo {year} {2004})},\ \Eprint
  {http://arxiv.org/abs/hep-lat/0311018} {arXiv:hep-lat/0311018} \BibitemShut
  {NoStop}%
\bibitem [{\citenamefont {Hasenfratz}\ and\ \citenamefont
  {Knechtli}(2001)}]{Hasenfratz:2001hp}%
  \BibitemOpen
  \bibfield  {author} {\bibinfo {author} {\bibfnamefont {A.}~\bibnamefont
  {Hasenfratz}}\ and\ \bibinfo {author} {\bibfnamefont {F.}~\bibnamefont
  {Knechtli}},\ }\href {\doibase 10.1103/PhysRevD.64.034504} {\bibfield
  {journal} {\bibinfo  {journal} {Phys. Rev. D}\ }\textbf {\bibinfo {volume}
  {64}},\ \bibinfo {pages} {034504} (\bibinfo {year} {2001})},\ \Eprint
  {http://arxiv.org/abs/hep-lat/0103029} {arXiv:hep-lat/0103029} \BibitemShut
  {NoStop}%
\bibitem [{\citenamefont {Capitani}\ \emph {et~al.}(2006)\citenamefont
  {Capitani}, \citenamefont {Durr},\ and\ \citenamefont
  {Hoelbling}}]{Capitani:2006ni}%
  \BibitemOpen
  \bibfield  {author} {\bibinfo {author} {\bibfnamefont {S.}~\bibnamefont
  {Capitani}}, \bibinfo {author} {\bibfnamefont {S.}~\bibnamefont {Durr}}, \
  and\ \bibinfo {author} {\bibfnamefont {C.}~\bibnamefont {Hoelbling}},\ }\href
  {\doibase 10.1088/1126-6708/2006/11/028} {\bibfield  {journal} {\bibinfo
  {journal} {JHEP}\ }\textbf {\bibinfo {volume} {11}},\ \bibinfo {pages} {028}
  (\bibinfo {year} {2006})},\ \Eprint {http://arxiv.org/abs/hep-lat/0607006}
  {arXiv:hep-lat/0607006} \BibitemShut {NoStop}%
\bibitem [{\citenamefont {Gusken}\ \emph {et~al.}(1989)\citenamefont {Gusken},
  \citenamefont {Low}, \citenamefont {Mutter}, \citenamefont {Sommer},
  \citenamefont {Patel},\ and\ \citenamefont {Schilling}}]{Gusken:1989ad}%
  \BibitemOpen
  \bibfield  {author} {\bibinfo {author} {\bibfnamefont {S.}~\bibnamefont
  {Gusken}}, \bibinfo {author} {\bibfnamefont {U.}~\bibnamefont {Low}},
  \bibinfo {author} {\bibfnamefont {K.~H.}\ \bibnamefont {Mutter}}, \bibinfo
  {author} {\bibfnamefont {R.}~\bibnamefont {Sommer}}, \bibinfo {author}
  {\bibfnamefont {A.}~\bibnamefont {Patel}}, \ and\ \bibinfo {author}
  {\bibfnamefont {K.}~\bibnamefont {Schilling}},\ }\href {\doibase
  10.1016/S0370-2693(89)80034-6} {\bibfield  {journal} {\bibinfo  {journal}
  {Phys. Lett. B}\ }\textbf {\bibinfo {volume} {227}},\ \bibinfo {pages} {266}
  (\bibinfo {year} {1989})}\BibitemShut {NoStop}%
\bibitem [{\citenamefont {Gusken}(1990)}]{Gusken:1989qx}%
  \BibitemOpen
  \bibfield  {author} {\bibinfo {author} {\bibfnamefont {S.}~\bibnamefont
  {Gusken}},\ }\href {\doibase 10.1016/0920-5632(90)90273-W} {\bibfield
  {journal} {\bibinfo  {journal} {Nucl. Phys. B Proc. Suppl.}\ }\textbf
  {\bibinfo {volume} {17}},\ \bibinfo {pages} {361} (\bibinfo {year}
  {1990})}\BibitemShut {NoStop}%
\bibitem [{\citenamefont {Alexandrou}\ \emph {et~al.}(1991)\citenamefont
  {Alexandrou}, \citenamefont {Jegerlehner}, \citenamefont {Gusken},
  \citenamefont {Schilling},\ and\ \citenamefont {Sommer}}]{Alexandrou:1990dq}%
  \BibitemOpen
  \bibfield  {author} {\bibinfo {author} {\bibfnamefont {C.}~\bibnamefont
  {Alexandrou}}, \bibinfo {author} {\bibfnamefont {F.}~\bibnamefont
  {Jegerlehner}}, \bibinfo {author} {\bibfnamefont {S.}~\bibnamefont {Gusken}},
  \bibinfo {author} {\bibfnamefont {K.}~\bibnamefont {Schilling}}, \ and\
  \bibinfo {author} {\bibfnamefont {R.}~\bibnamefont {Sommer}},\ }\href
  {\doibase 10.1016/0370-2693(91)90219-G} {\bibfield  {journal} {\bibinfo
  {journal} {Phys. Lett. B}\ }\textbf {\bibinfo {volume} {256}},\ \bibinfo
  {pages} {60} (\bibinfo {year} {1991})}\BibitemShut {NoStop}%
\bibitem [{\citenamefont {L\"uscher}(2010)}]{Luscher:2010iy}%
  \BibitemOpen
  \bibfield  {author} {\bibinfo {author} {\bibfnamefont {M.}~\bibnamefont
  {L\"uscher}},\ }\href {\doibase 10.1007/JHEP08(2010)071} {\bibfield
  {journal} {\bibinfo  {journal} {JHEP}\ }\textbf {\bibinfo {volume} {08}},\
  \bibinfo {pages} {071} (\bibinfo {year} {2010})},\ \bibinfo {note} {[Erratum:
  JHEP 03, 092 (2014)]},\ \Eprint {http://arxiv.org/abs/1006.4518}
  {arXiv:1006.4518 [hep-lat]} \BibitemShut {NoStop}%
\bibitem [{\citenamefont {Bali}\ \emph {et~al.}(2016)\citenamefont {Bali},
  \citenamefont {Lang}, \citenamefont {Musch},\ and\ \citenamefont
  {Sch\"afer}}]{Bali:2016lva}%
  \BibitemOpen
  \bibfield  {author} {\bibinfo {author} {\bibfnamefont {G.~S.}\ \bibnamefont
  {Bali}}, \bibinfo {author} {\bibfnamefont {B.}~\bibnamefont {Lang}}, \bibinfo
  {author} {\bibfnamefont {B.~U.}\ \bibnamefont {Musch}}, \ and\ \bibinfo
  {author} {\bibfnamefont {A.}~\bibnamefont {Sch\"afer}},\ }\href {\doibase
  10.1103/PhysRevD.93.094515} {\bibfield  {journal} {\bibinfo  {journal} {Phys.
  Rev. D}\ }\textbf {\bibinfo {volume} {93}},\ \bibinfo {pages} {094515}
  (\bibinfo {year} {2016})},\ \Eprint {http://arxiv.org/abs/1602.05525}
  {arXiv:1602.05525 [hep-lat]} \BibitemShut {NoStop}%
\bibitem [{\citenamefont {Fox}\ \emph {et~al.}(1982)\citenamefont {Fox},
  \citenamefont {Gupta}, \citenamefont {Martin},\ and\ \citenamefont
  {Otto}}]{Fox:1981xz}%
  \BibitemOpen
  \bibfield  {author} {\bibinfo {author} {\bibfnamefont {G.}~\bibnamefont
  {Fox}}, \bibinfo {author} {\bibfnamefont {R.}~\bibnamefont {Gupta}}, \bibinfo
  {author} {\bibfnamefont {O.}~\bibnamefont {Martin}}, \ and\ \bibinfo {author}
  {\bibfnamefont {S.}~\bibnamefont {Otto}},\ }\href {\doibase
  10.1016/0550-3213(82)90384-4} {\bibfield  {journal} {\bibinfo  {journal}
  {Nucl. Phys. B}\ }\textbf {\bibinfo {volume} {205}},\ \bibinfo {pages} {188}
  (\bibinfo {year} {1982})}\BibitemShut {NoStop}%
\bibitem [{\citenamefont {Luscher}\ and\ \citenamefont
  {Wolff}(1990)}]{Luscher:1990ck}%
  \BibitemOpen
  \bibfield  {author} {\bibinfo {author} {\bibfnamefont {M.}~\bibnamefont
  {Luscher}}\ and\ \bibinfo {author} {\bibfnamefont {U.}~\bibnamefont
  {Wolff}},\ }\href {\doibase 10.1016/0550-3213(90)90540-T} {\bibfield
  {journal} {\bibinfo  {journal} {Nucl. Phys. B}\ }\textbf {\bibinfo {volume}
  {339}},\ \bibinfo {pages} {222} (\bibinfo {year} {1990})}\BibitemShut
  {NoStop}%
\bibitem [{\citenamefont {Michael}\ and\ \citenamefont
  {Teasdale}(1983)}]{Michael:1982gb}%
  \BibitemOpen
  \bibfield  {author} {\bibinfo {author} {\bibfnamefont {C.}~\bibnamefont
  {Michael}}\ and\ \bibinfo {author} {\bibfnamefont {I.}~\bibnamefont
  {Teasdale}},\ }\href {\doibase 10.1016/0550-3213(83)90674-0} {\bibfield
  {journal} {\bibinfo  {journal} {Nucl. Phys. B}\ }\textbf {\bibinfo {volume}
  {215}},\ \bibinfo {pages} {433} (\bibinfo {year} {1983})}\BibitemShut
  {NoStop}%
\bibitem [{\citenamefont {Michael}(1985)}]{Michael:1985ne}%
  \BibitemOpen
  \bibfield  {author} {\bibinfo {author} {\bibfnamefont {C.}~\bibnamefont
  {Michael}},\ }\href {\doibase 10.1016/0550-3213(85)90297-4} {\bibfield
  {journal} {\bibinfo  {journal} {Nucl. Phys. B}\ }\textbf {\bibinfo {volume}
  {259}},\ \bibinfo {pages} {58} (\bibinfo {year} {1985})}\BibitemShut
  {NoStop}%
\bibitem [{\citenamefont {Blossier}\ \emph {et~al.}(2008)\citenamefont
  {Blossier}, \citenamefont {von Hippel}, \citenamefont {Mendes}, \citenamefont
  {Sommer},\ and\ \citenamefont {Della~Morte}}]{Blossier:2008tx}%
  \BibitemOpen
  \bibfield  {author} {\bibinfo {author} {\bibfnamefont {B.}~\bibnamefont
  {Blossier}}, \bibinfo {author} {\bibfnamefont {G.}~\bibnamefont {von
  Hippel}}, \bibinfo {author} {\bibfnamefont {T.}~\bibnamefont {Mendes}},
  \bibinfo {author} {\bibfnamefont {R.}~\bibnamefont {Sommer}}, \ and\ \bibinfo
  {author} {\bibfnamefont {M.}~\bibnamefont {Della~Morte}},\ }\href {\doibase
  10.22323/1.066.0135} {\bibfield  {journal} {\bibinfo  {journal} {PoS}\
  }\textbf {\bibinfo {volume} {LATTICE2008}},\ \bibinfo {pages} {135} (\bibinfo
  {year} {2008})},\ \Eprint {http://arxiv.org/abs/0808.1017} {arXiv:0808.1017
  [hep-lat]} \BibitemShut {NoStop}%
\bibitem [{\citenamefont {Blossier}\ \emph {et~al.}(2009)\citenamefont
  {Blossier}, \citenamefont {Della~Morte}, \citenamefont {von Hippel},
  \citenamefont {Mendes},\ and\ \citenamefont {Sommer}}]{Blossier:2009kd}%
  \BibitemOpen
  \bibfield  {author} {\bibinfo {author} {\bibfnamefont {B.}~\bibnamefont
  {Blossier}}, \bibinfo {author} {\bibfnamefont {M.}~\bibnamefont
  {Della~Morte}}, \bibinfo {author} {\bibfnamefont {G.}~\bibnamefont {von
  Hippel}}, \bibinfo {author} {\bibfnamefont {T.}~\bibnamefont {Mendes}}, \
  and\ \bibinfo {author} {\bibfnamefont {R.}~\bibnamefont {Sommer}},\ }\href
  {\doibase 10.1088/1126-6708/2009/04/094} {\bibfield  {journal} {\bibinfo
  {journal} {JHEP}\ }\textbf {\bibinfo {volume} {04}},\ \bibinfo {pages} {094}
  (\bibinfo {year} {2009})},\ \Eprint {http://arxiv.org/abs/0902.1265}
  {arXiv:0902.1265 [hep-lat]} \BibitemShut {NoStop}%
\bibitem [{\citenamefont {Mahbub}\ \emph {et~al.}(2009)\citenamefont {Mahbub},
  \citenamefont {O.~Cais}, \citenamefont {Kamleh}, \citenamefont {Lasscock},
  \citenamefont {Leinweber},\ and\ \citenamefont {Williams}}]{Mahbub:2009nr}%
  \BibitemOpen
  \bibfield  {author} {\bibinfo {author} {\bibfnamefont {M.~S.}\ \bibnamefont
  {Mahbub}}, \bibinfo {author} {\bibfnamefont {A.}~\bibnamefont {O.~Cais}},
  \bibinfo {author} {\bibfnamefont {W.}~\bibnamefont {Kamleh}}, \bibinfo
  {author} {\bibfnamefont {B.~G.}\ \bibnamefont {Lasscock}}, \bibinfo {author}
  {\bibfnamefont {D.~B.}\ \bibnamefont {Leinweber}}, \ and\ \bibinfo {author}
  {\bibfnamefont {A.~G.}\ \bibnamefont {Williams}},\ }\href {\doibase
  10.1103/PhysRevD.80.054507} {\bibfield  {journal} {\bibinfo  {journal} {Phys.
  Rev. D}\ }\textbf {\bibinfo {volume} {80}},\ \bibinfo {pages} {054507}
  (\bibinfo {year} {2009})},\ \Eprint {http://arxiv.org/abs/0905.3616}
  {arXiv:0905.3616 [hep-lat]} \BibitemShut {NoStop}%
\bibitem [{\citenamefont {Detmold}\ and\ \citenamefont
  {Endres}(2015)}]{Detmold:2014rfa}%
  \BibitemOpen
  \bibfield  {author} {\bibinfo {author} {\bibfnamefont {W.}~\bibnamefont
  {Detmold}}\ and\ \bibinfo {author} {\bibfnamefont {M.~G.}\ \bibnamefont
  {Endres}},\ }\href {\doibase 10.22323/1.214.0170} {\bibfield  {journal}
  {\bibinfo  {journal} {PoS}\ }\textbf {\bibinfo {volume} {LATTICE2014}},\
  \bibinfo {pages} {170} (\bibinfo {year} {2015})},\ \Eprint
  {http://arxiv.org/abs/1409.5667} {arXiv:1409.5667 [hep-lat]} \BibitemShut
  {NoStop}%
\bibitem [{\citenamefont {Detmold}\ and\ \citenamefont
  {Endres}(2014)}]{Detmold:2014hla}%
  \BibitemOpen
  \bibfield  {author} {\bibinfo {author} {\bibfnamefont {W.}~\bibnamefont
  {Detmold}}\ and\ \bibinfo {author} {\bibfnamefont {M.~G.}\ \bibnamefont
  {Endres}},\ }\href {\doibase 10.1103/PhysRevD.90.034503} {\bibfield
  {journal} {\bibinfo  {journal} {Phys. Rev. D}\ }\textbf {\bibinfo {volume}
  {90}},\ \bibinfo {pages} {034503} (\bibinfo {year} {2014})},\ \Eprint
  {http://arxiv.org/abs/1404.6816} {arXiv:1404.6816 [hep-lat]} \BibitemShut
  {NoStop}%
\bibitem [{\citenamefont {Fleming}(2004)}]{Fleming:2004hs}%
  \BibitemOpen
  \bibfield  {author} {\bibinfo {author} {\bibfnamefont {G.~T.}\ \bibnamefont
  {Fleming}},\ }in\ \href@noop {} {\emph {\bibinfo {booktitle} {{3rd
  International Workshop on Numerical Analysis and Lattice QCD}}}}\ (\bibinfo
  {year} {2004})\ pp.\ \bibinfo {pages} {143--152},\ \Eprint
  {http://arxiv.org/abs/hep-lat/0403023} {arXiv:hep-lat/0403023} \BibitemShut
  {NoStop}%
\bibitem [{\citenamefont {Lin}\ and\ \citenamefont {Cohen}(2007)}]{Lin:2007iq}%
  \BibitemOpen
  \bibfield  {author} {\bibinfo {author} {\bibfnamefont {H.-W.}\ \bibnamefont
  {Lin}}\ and\ \bibinfo {author} {\bibfnamefont {S.~D.}\ \bibnamefont
  {Cohen}},\ }in\ \href@noop {} {\emph {\bibinfo {booktitle} {{4th
  International Workshop on Numerical Analysis and Lattice QCD}}}}\ (\bibinfo
  {year} {2007})\ \Eprint {http://arxiv.org/abs/0709.1902} {arXiv:0709.1902
  [hep-lat]} \BibitemShut {NoStop}%
\bibitem [{\citenamefont {Beane}\ \emph
  {et~al.}(2009{\natexlab{a}})\citenamefont {Beane}, \citenamefont {Detmold},
  \citenamefont {Luu}, \citenamefont {Orginos}, \citenamefont {Parreno},
  \citenamefont {Savage}, \citenamefont {Torok},\ and\ \citenamefont
  {Walker-Loud}}]{Beane:2009kya}%
  \BibitemOpen
  \bibfield  {author} {\bibinfo {author} {\bibfnamefont {S.~R.}\ \bibnamefont
  {Beane}}, \bibinfo {author} {\bibfnamefont {W.}~\bibnamefont {Detmold}},
  \bibinfo {author} {\bibfnamefont {T.~C.}\ \bibnamefont {Luu}}, \bibinfo
  {author} {\bibfnamefont {K.}~\bibnamefont {Orginos}}, \bibinfo {author}
  {\bibfnamefont {A.}~\bibnamefont {Parreno}}, \bibinfo {author} {\bibfnamefont
  {M.~J.}\ \bibnamefont {Savage}}, \bibinfo {author} {\bibfnamefont
  {A.}~\bibnamefont {Torok}}, \ and\ \bibinfo {author} {\bibfnamefont
  {A.}~\bibnamefont {Walker-Loud}},\ }\href {\doibase
  10.1103/PhysRevD.79.114502} {\bibfield  {journal} {\bibinfo  {journal} {Phys.
  Rev. D}\ }\textbf {\bibinfo {volume} {79}},\ \bibinfo {pages} {114502}
  (\bibinfo {year} {2009}{\natexlab{a}})},\ \Eprint
  {http://arxiv.org/abs/0903.2990} {arXiv:0903.2990 [hep-lat]} \BibitemShut
  {NoStop}%
\bibitem [{\citenamefont {Beane}\ \emph
  {et~al.}(2009{\natexlab{b}})\citenamefont {Beane}, \citenamefont {Detmold},
  \citenamefont {Luu}, \citenamefont {Orginos}, \citenamefont {Parreno},
  \citenamefont {Savage}, \citenamefont {Torok},\ and\ \citenamefont
  {Walker-Loud}}]{Beane:2009gs}%
  \BibitemOpen
  \bibfield  {author} {\bibinfo {author} {\bibfnamefont {S.~R.}\ \bibnamefont
  {Beane}}, \bibinfo {author} {\bibfnamefont {W.}~\bibnamefont {Detmold}},
  \bibinfo {author} {\bibfnamefont {T.~C.}\ \bibnamefont {Luu}}, \bibinfo
  {author} {\bibfnamefont {K.}~\bibnamefont {Orginos}}, \bibinfo {author}
  {\bibfnamefont {A.}~\bibnamefont {Parreno}}, \bibinfo {author} {\bibfnamefont
  {M.~J.}\ \bibnamefont {Savage}}, \bibinfo {author} {\bibfnamefont
  {A.}~\bibnamefont {Torok}}, \ and\ \bibinfo {author} {\bibfnamefont
  {A.}~\bibnamefont {Walker-Loud}},\ }\href {\doibase
  10.1103/PhysRevD.80.074501} {\bibfield  {journal} {\bibinfo  {journal} {Phys.
  Rev. D}\ }\textbf {\bibinfo {volume} {80}},\ \bibinfo {pages} {074501}
  (\bibinfo {year} {2009}{\natexlab{b}})},\ \Eprint
  {http://arxiv.org/abs/0905.0466} {arXiv:0905.0466 [hep-lat]} \BibitemShut
  {NoStop}%
\bibitem [{\citenamefont {Schollw\"{o}ck}(2011)}]{Schollwoeck2011}%
  \BibitemOpen
  \bibfield  {author} {\bibinfo {author} {\bibfnamefont {U.}~\bibnamefont
  {Schollw\"{o}ck}},\ }\href {\doibase 10.1016/j.aop.2010.09.012} {\bibfield
  {journal} {\bibinfo  {journal} {Ann. Phys.}\ }\textbf {\bibinfo {volume}
  {326}},\ \bibinfo {pages} {96} (\bibinfo {year} {2011})}\BibitemShut
  {NoStop}%
\bibitem [{\citenamefont {Or{\'u}s}(2014)}]{Orus2014a}%
  \BibitemOpen
  \bibfield  {author} {\bibinfo {author} {\bibfnamefont {R.}~\bibnamefont
  {Or{\'u}s}},\ }\href {\doibase 10.1016/j.aop.2014.06.013} {\bibfield
  {journal} {\bibinfo  {journal} {Ann. Phys.}\ }\textbf {\bibinfo {volume}
  {349}},\ \bibinfo {pages} {117 } (\bibinfo {year} {2014})}\BibitemShut
  {NoStop}%
\bibitem [{\citenamefont {Schwinger}(1962)}]{Schwinger1962}%
  \BibitemOpen
  \bibfield  {author} {\bibinfo {author} {\bibfnamefont {J.}~\bibnamefont
  {Schwinger}},\ }\href {\doibase 10.1103/PhysRev.128.2425} {\bibfield
  {journal} {\bibinfo  {journal} {Phys. Rev.}\ }\textbf {\bibinfo {volume}
  {128}},\ \bibinfo {pages} {2425} (\bibinfo {year} {1962})}\BibitemShut
  {NoStop}%
\bibitem [{\citenamefont {Coleman}\ \emph {et~al.}(1975)\citenamefont
  {Coleman}, \citenamefont {Jackiw},\ and\ \citenamefont
  {Susskind}}]{Coleman1975}%
  \BibitemOpen
  \bibfield  {author} {\bibinfo {author} {\bibfnamefont {S.}~\bibnamefont
  {Coleman}}, \bibinfo {author} {\bibfnamefont {R.}~\bibnamefont {Jackiw}}, \
  and\ \bibinfo {author} {\bibfnamefont {L.}~\bibnamefont {Susskind}},\ }\href
  {\doibase 10.1016/0003-4916(75)90212-2} {\bibfield  {journal} {\bibinfo
  {journal} {Ann. Phys.}\ }\textbf {\bibinfo {volume} {93}},\ \bibinfo {pages}
  {267} (\bibinfo {year} {1975})}\BibitemShut {NoStop}%
\bibitem [{\citenamefont {Coleman}(1976{\natexlab{a}})}]{Coleman1976}%
  \BibitemOpen
  \bibfield  {author} {\bibinfo {author} {\bibfnamefont {S.}~\bibnamefont
  {Coleman}},\ }\href
  {http://www.sciencedirect.com/science/article/pii/0003491676902803}
  {\bibfield  {journal} {\bibinfo  {journal} {Ann. Phys. (N.Y.)}\ }\textbf
  {\bibinfo {volume} {101}},\ \bibinfo {pages} {239} (\bibinfo {year}
  {1976}{\natexlab{a}})}\BibitemShut {NoStop}%
\bibitem [{\citenamefont {Zache}\ \emph {et~al.}(2018)\citenamefont {Zache},
  \citenamefont {Hebenstreit}, \citenamefont {Jendrzejewski}, \citenamefont
  {Oberthaler}, \citenamefont {Berges},\ and\ \citenamefont
  {Hauke}}]{Zache2018}%
  \BibitemOpen
  \bibfield  {author} {\bibinfo {author} {\bibfnamefont {T.~V.}\ \bibnamefont
  {Zache}}, \bibinfo {author} {\bibfnamefont {F.}~\bibnamefont {Hebenstreit}},
  \bibinfo {author} {\bibfnamefont {F.}~\bibnamefont {Jendrzejewski}}, \bibinfo
  {author} {\bibfnamefont {M.~K.}\ \bibnamefont {Oberthaler}}, \bibinfo
  {author} {\bibfnamefont {J.}~\bibnamefont {Berges}}, \ and\ \bibinfo {author}
  {\bibfnamefont {P.}~\bibnamefont {Hauke}},\ }\href {\doibase
  10.1088/2058-9565/aac33b} {\bibfield  {journal} {\bibinfo  {journal} {Quantum
  Science and Technology}\ }\textbf {\bibinfo {volume} {3}},\ \bibinfo {pages}
  {034010} (\bibinfo {year} {2018})}\BibitemShut {NoStop}%
\bibitem [{\citenamefont {Angelides}\ \emph {et~al.}(2023)\citenamefont
  {Angelides}, \citenamefont {Funcke}, \citenamefont {Jansen},\ and\
  \citenamefont {Kühn}}]{Angelides2023}%
  \BibitemOpen
  \bibfield  {author} {\bibinfo {author} {\bibfnamefont {T.}~\bibnamefont
  {Angelides}}, \bibinfo {author} {\bibfnamefont {L.}~\bibnamefont {Funcke}},
  \bibinfo {author} {\bibfnamefont {K.}~\bibnamefont {Jansen}}, \ and\ \bibinfo
  {author} {\bibfnamefont {S.}~\bibnamefont {Kühn}},\ }\href {\doibase
  10.1103/physrevd.108.014516} {\bibfield  {journal} {\bibinfo  {journal}
  {Physical Review D}\ }\textbf {\bibinfo {volume} {108}},\ \bibinfo {pages}
  {014516} (\bibinfo {year} {2023})}\BibitemShut {NoStop}%
\bibitem [{\citenamefont {Coleman}(1976{\natexlab{b}})}]{Coleman:1976uz}%
  \BibitemOpen
  \bibfield  {author} {\bibinfo {author} {\bibfnamefont {S.~R.}\ \bibnamefont
  {Coleman}},\ }\href {\doibase 10.1016/0003-4916(76)90280-3} {\bibfield
  {journal} {\bibinfo  {journal} {Annals Phys.}\ }\textbf {\bibinfo {volume}
  {101}},\ \bibinfo {pages} {239} (\bibinfo {year}
  {1976}{\natexlab{b}})}\BibitemShut {NoStop}%
\bibitem [{\citenamefont {Susskind}(1977)}]{Susskind:1976jm}%
  \BibitemOpen
  \bibfield  {author} {\bibinfo {author} {\bibfnamefont {L.}~\bibnamefont
  {Susskind}},\ }\href {\doibase 10.1103/PhysRevD.16.3031} {\bibfield
  {journal} {\bibinfo  {journal} {Phys. Rev. D}\ }\textbf {\bibinfo {volume}
  {16}},\ \bibinfo {pages} {3031} (\bibinfo {year} {1977})}\BibitemShut
  {NoStop}%
\bibitem [{\citenamefont {Sharatchandra}\ \emph {et~al.}(1981)\citenamefont
  {Sharatchandra}, \citenamefont {Thun},\ and\ \citenamefont
  {Weisz}}]{Sharatchandra:1981si}%
  \BibitemOpen
  \bibfield  {author} {\bibinfo {author} {\bibfnamefont {H.~S.}\ \bibnamefont
  {Sharatchandra}}, \bibinfo {author} {\bibfnamefont {H.~J.}\ \bibnamefont
  {Thun}}, \ and\ \bibinfo {author} {\bibfnamefont {P.}~\bibnamefont {Weisz}},\
  }\href {\doibase 10.1016/0550-3213(81)90200-5} {\bibfield  {journal}
  {\bibinfo  {journal} {Nucl. Phys. B}\ }\textbf {\bibinfo {volume} {192}},\
  \bibinfo {pages} {205} (\bibinfo {year} {1981})}\BibitemShut {NoStop}%
\bibitem [{\citenamefont {Bridgeman}\ and\ \citenamefont
  {Chubb}(2017)}]{Bridgeman2017}%
  \BibitemOpen
  \bibfield  {author} {\bibinfo {author} {\bibfnamefont {J.~C.}\ \bibnamefont
  {Bridgeman}}\ and\ \bibinfo {author} {\bibfnamefont {C.~T.}\ \bibnamefont
  {Chubb}},\ }\href {\doibase 10.1088/1751-8121/aa6dc3} {\bibfield  {journal}
  {\bibinfo  {journal} {J. Phys. A: Math. Theor.}\ }\textbf {\bibinfo {volume}
  {50}},\ \bibinfo {pages} {223001} (\bibinfo {year} {2017})}\BibitemShut
  {NoStop}%
\bibitem [{\citenamefont {Ba\~{n}uls}\ \emph {et~al.}(2013)\citenamefont
  {Ba\~{n}uls}, \citenamefont {Cichy}, \citenamefont {Jansen},\ and\
  \citenamefont {Cirac}}]{Banuls2013}%
  \BibitemOpen
  \bibfield  {author} {\bibinfo {author} {\bibfnamefont {M.~C.}\ \bibnamefont
  {Ba\~{n}uls}}, \bibinfo {author} {\bibfnamefont {K.}~\bibnamefont {Cichy}},
  \bibinfo {author} {\bibfnamefont {K.}~\bibnamefont {Jansen}}, \ and\ \bibinfo
  {author} {\bibfnamefont {J.~I.}\ \bibnamefont {Cirac}},\ }\href {\doibase
  10.1007/JHEP11(2013)158} {\bibfield  {journal} {\bibinfo  {journal} {J. High
  Energy Phys.}\ }\textbf {\bibinfo {volume} {2013}},\ \bibinfo {pages} {158}
  (\bibinfo {year} {2013})}\BibitemShut {NoStop}%
\bibitem [{\citenamefont {Kuberski}(2024)}]{Kuberski:2023zky}%
  \BibitemOpen
  \bibfield  {author} {\bibinfo {author} {\bibfnamefont {S.}~\bibnamefont
  {Kuberski}},\ }\href {\doibase 10.1016/j.cpc.2024.109173} {\bibfield
  {journal} {\bibinfo  {journal} {Comput. Phys. Commun.}\ }\textbf {\bibinfo
  {volume} {300}},\ \bibinfo {pages} {109173} (\bibinfo {year} {2024})},\
  \Eprint {http://arxiv.org/abs/2306.02385} {arXiv:2306.02385 [hep-lat]}
  \BibitemShut {NoStop}%
\bibitem [{\citenamefont {Kogut}\ \emph {et~al.}(1979)\citenamefont {Kogut},
  \citenamefont {Pearson},\ and\ \citenamefont {Shigemitsu}}]{Kogut:1979vg}%
  \BibitemOpen
  \bibfield  {author} {\bibinfo {author} {\bibfnamefont {J.~B.}\ \bibnamefont
  {Kogut}}, \bibinfo {author} {\bibfnamefont {R.~B.}\ \bibnamefont {Pearson}},
  \ and\ \bibinfo {author} {\bibfnamefont {J.}~\bibnamefont {Shigemitsu}},\
  }\href {\doibase 10.1103/PhysRevLett.43.484} {\bibfield  {journal} {\bibinfo
  {journal} {Phys. Rev. Lett.}\ }\textbf {\bibinfo {volume} {43}},\ \bibinfo
  {pages} {484} (\bibinfo {year} {1979})}\BibitemShut {NoStop}%
\bibitem [{\citenamefont {Dashen}\ and\ \citenamefont
  {Gross}(1981)}]{Dashen:1980vm}%
  \BibitemOpen
  \bibfield  {author} {\bibinfo {author} {\bibfnamefont {R.~F.}\ \bibnamefont
  {Dashen}}\ and\ \bibinfo {author} {\bibfnamefont {D.~J.}\ \bibnamefont
  {Gross}},\ }\href {\doibase 10.1103/PhysRevD.23.2340} {\bibfield  {journal}
  {\bibinfo  {journal} {Phys. Rev. D}\ }\textbf {\bibinfo {volume} {23}},\
  \bibinfo {pages} {2340} (\bibinfo {year} {1981})}\BibitemShut {NoStop}%
\bibitem [{\citenamefont {Hasenfratz}\ and\ \citenamefont
  {Hasenfratz}(1981)}]{Hasenfratz:1981tw}%
  \BibitemOpen
  \bibfield  {author} {\bibinfo {author} {\bibfnamefont {A.}~\bibnamefont
  {Hasenfratz}}\ and\ \bibinfo {author} {\bibfnamefont {P.}~\bibnamefont
  {Hasenfratz}},\ }\href {\doibase 10.1016/0550-3213(81)90526-5} {\bibfield
  {journal} {\bibinfo  {journal} {Nucl. Phys. B}\ }\textbf {\bibinfo {volume}
  {193}},\ \bibinfo {pages} {210} (\bibinfo {year} {1981})}\BibitemShut
  {NoStop}%
\bibitem [{\citenamefont {Karsch}(1982)}]{Karsch:1982ve}%
  \BibitemOpen
  \bibfield  {author} {\bibinfo {author} {\bibfnamefont {F.}~\bibnamefont
  {Karsch}},\ }\href {\doibase 10.1016/0550-3213(82)90390-X} {\bibfield
  {journal} {\bibinfo  {journal} {Nucl. Phys. B}\ }\textbf {\bibinfo {volume}
  {205}},\ \bibinfo {pages} {285} (\bibinfo {year} {1982})}\BibitemShut
  {NoStop}%
\bibitem [{\citenamefont {Loan}\ \emph {et~al.}(2003)\citenamefont {Loan},
  \citenamefont {Brunner}, \citenamefont {Sloggett},\ and\ \citenamefont
  {Hamer}}]{Loan:2002ej}%
  \BibitemOpen
  \bibfield  {author} {\bibinfo {author} {\bibfnamefont {M.}~\bibnamefont
  {Loan}}, \bibinfo {author} {\bibfnamefont {M.}~\bibnamefont {Brunner}},
  \bibinfo {author} {\bibfnamefont {C.}~\bibnamefont {Sloggett}}, \ and\
  \bibinfo {author} {\bibfnamefont {C.}~\bibnamefont {Hamer}},\ }\href
  {\doibase 10.1103/PhysRevD.68.034504} {\bibfield  {journal} {\bibinfo
  {journal} {Phys. Rev. D}\ }\textbf {\bibinfo {volume} {68}},\ \bibinfo
  {pages} {034504} (\bibinfo {year} {2003})},\ \Eprint
  {http://arxiv.org/abs/hep-lat/0209159} {arXiv:hep-lat/0209159} \BibitemShut
  {NoStop}%
\bibitem [{\citenamefont {Byrnes}\ \emph {et~al.}(2004)\citenamefont {Byrnes},
  \citenamefont {Loan}, \citenamefont {Hamer}, \citenamefont {Bonnet},
  \citenamefont {Leinweber}, \citenamefont {Williams},\ and\ \citenamefont
  {Zanotti}}]{Byrnes:2003gg}%
  \BibitemOpen
  \bibfield  {author} {\bibinfo {author} {\bibfnamefont {T.~M.~R.}\
  \bibnamefont {Byrnes}}, \bibinfo {author} {\bibfnamefont {M.}~\bibnamefont
  {Loan}}, \bibinfo {author} {\bibfnamefont {C.~J.}\ \bibnamefont {Hamer}},
  \bibinfo {author} {\bibfnamefont {F.~D.~R.}\ \bibnamefont {Bonnet}}, \bibinfo
  {author} {\bibfnamefont {D.~B.}\ \bibnamefont {Leinweber}}, \bibinfo {author}
  {\bibfnamefont {A.~G.}\ \bibnamefont {Williams}}, \ and\ \bibinfo {author}
  {\bibfnamefont {J.~M.}\ \bibnamefont {Zanotti}},\ }\href {\doibase
  10.1103/PhysRevD.69.074509} {\bibfield  {journal} {\bibinfo  {journal} {Phys.
  Rev. D}\ }\textbf {\bibinfo {volume} {69}},\ \bibinfo {pages} {074509}
  (\bibinfo {year} {2004})},\ \Eprint {http://arxiv.org/abs/hep-lat/0311014}
  {arXiv:hep-lat/0311014} \BibitemShut {NoStop}%
\bibitem [{\citenamefont {Funcke}\ \emph
  {et~al.}(2023{\natexlab{c}})\citenamefont {Funcke}, \citenamefont {Gro\ss{}},
  \citenamefont {Jansen}, \citenamefont {K\"uhn}, \citenamefont {Romiti},\ and\
  \citenamefont {Urbach}}]{Funcke:2022opx}%
  \BibitemOpen
  \bibfield  {author} {\bibinfo {author} {\bibfnamefont {L.}~\bibnamefont
  {Funcke}}, \bibinfo {author} {\bibfnamefont {C.~F.}\ \bibnamefont
  {Gro\ss{}}}, \bibinfo {author} {\bibfnamefont {K.}~\bibnamefont {Jansen}},
  \bibinfo {author} {\bibfnamefont {S.}~\bibnamefont {K\"uhn}}, \bibinfo
  {author} {\bibfnamefont {S.}~\bibnamefont {Romiti}}, \ and\ \bibinfo {author}
  {\bibfnamefont {C.}~\bibnamefont {Urbach}},\ }\href {\doibase
  10.22323/1.430.0292} {\bibfield  {journal} {\bibinfo  {journal} {PoS}\
  }\textbf {\bibinfo {volume} {LATTICE2022}},\ \bibinfo {pages} {292} (\bibinfo
  {year} {2023}{\natexlab{c}})},\ \Eprint {http://arxiv.org/abs/2212.09627}
  {arXiv:2212.09627 [hep-lat]} \BibitemShut {NoStop}%
\bibitem [{\citenamefont {Della~Morte}\ \emph {et~al.}(2012)\citenamefont
  {Della~Morte}, \citenamefont {Jaeger}, \citenamefont {Rae},\ and\
  \citenamefont {Wittig}}]{DellaMorte:2012xc}%
  \BibitemOpen
  \bibfield  {author} {\bibinfo {author} {\bibfnamefont {M.}~\bibnamefont
  {Della~Morte}}, \bibinfo {author} {\bibfnamefont {B.}~\bibnamefont {Jaeger}},
  \bibinfo {author} {\bibfnamefont {T.}~\bibnamefont {Rae}}, \ and\ \bibinfo
  {author} {\bibfnamefont {H.}~\bibnamefont {Wittig}},\ }\href {\doibase
  10.1140/epja/i2012-12139-9} {\bibfield  {journal} {\bibinfo  {journal} {Eur.
  Phys. J. A}\ }\textbf {\bibinfo {volume} {48}},\ \bibinfo {pages} {139}
  (\bibinfo {year} {2012})},\ \Eprint {http://arxiv.org/abs/1208.0189}
  {arXiv:1208.0189 [hep-lat]} \BibitemShut {NoStop}%
\bibitem [{\citenamefont {Wu}\ \emph {et~al.}(2018)\citenamefont {Wu},
  \citenamefont {Kamleh}, \citenamefont {Leinweber}, \citenamefont {Young},\
  and\ \citenamefont {Zanotti}}]{Wu:2018tvt}%
  \BibitemOpen
  \bibfield  {author} {\bibinfo {author} {\bibfnamefont {J.~J.}\ \bibnamefont
  {Wu}}, \bibinfo {author} {\bibfnamefont {W.}~\bibnamefont {Kamleh}}, \bibinfo
  {author} {\bibfnamefont {D.~t.}\ \bibnamefont {Leinweber}}, \bibinfo {author}
  {\bibfnamefont {R.~D.}\ \bibnamefont {Young}}, \ and\ \bibinfo {author}
  {\bibfnamefont {J.~M.}\ \bibnamefont {Zanotti}},\ }\href {\doibase
  10.1088/1361-6471/aaeb9e} {\bibfield  {journal} {\bibinfo  {journal} {J.
  Phys. G}\ }\textbf {\bibinfo {volume} {45}},\ \bibinfo {pages} {125102}
  (\bibinfo {year} {2018})},\ \Eprint {http://arxiv.org/abs/1807.09429}
  {arXiv:1807.09429 [hep-lat]} \BibitemShut {NoStop}%
\bibitem [{\citenamefont {Hastings}(2007)}]{Hastings2007}%
  \BibitemOpen
  \bibfield  {author} {\bibinfo {author} {\bibfnamefont {M.~B.}\ \bibnamefont
  {Hastings}},\ }\href {http://stacks.iop.org/1742-5468/2007/i=08/a=P08024}
  {\bibfield  {journal} {\bibinfo  {journal} {J. Stat. Mech.}\ }\textbf
  {\bibinfo {volume} {2007}},\ \bibinfo {pages} {P08024} (\bibinfo {year}
  {2007})}\BibitemShut {NoStop}%
\bibitem [{\citenamefont {Eisert}\ \emph {et~al.}(2010)\citenamefont {Eisert},
  \citenamefont {Cramer},\ and\ \citenamefont {Plenio}}]{Eisert2010}%
  \BibitemOpen
  \bibfield  {author} {\bibinfo {author} {\bibfnamefont {J.}~\bibnamefont
  {Eisert}}, \bibinfo {author} {\bibfnamefont {M.}~\bibnamefont {Cramer}}, \
  and\ \bibinfo {author} {\bibfnamefont {M.~B.}\ \bibnamefont {Plenio}},\
  }\href {\doibase 10.1103/RevModPhys.82.277} {\bibfield  {journal} {\bibinfo
  {journal} {Rev. Mod. Phys.}\ }\textbf {\bibinfo {volume} {82}},\ \bibinfo
  {pages} {277} (\bibinfo {year} {2010})}\BibitemShut {NoStop}%
\bibitem [{\citenamefont {Vidal}(2003)}]{Vidal2003}%
  \BibitemOpen
  \bibfield  {author} {\bibinfo {author} {\bibfnamefont {G.}~\bibnamefont
  {Vidal}},\ }\href {\doibase 10.1103/PhysRevLett.91.147902} {\bibfield
  {journal} {\bibinfo  {journal} {Phys. Rev. Lett.}\ }\textbf {\bibinfo
  {volume} {91}},\ \bibinfo {pages} {147902} (\bibinfo {year}
  {2003})}\BibitemShut {NoStop}%
\bibitem [{\citenamefont {Verstraete}\ and\ \citenamefont
  {Cirac}(2004)}]{Verstraete2004b}%
  \BibitemOpen
  \bibfield  {author} {\bibinfo {author} {\bibfnamefont {F.}~\bibnamefont
  {Verstraete}}\ and\ \bibinfo {author} {\bibfnamefont {J.~I.}\ \bibnamefont
  {Cirac}},\ }\href {http://arxiv.org/abs/cond-mat/0407066} {\bibfield
  {journal} {\bibinfo  {journal} {arXiv:cond-mat/0407066}\ } (\bibinfo {year}
  {2004})}\BibitemShut {NoStop}%
\bibitem [{\citenamefont {Shi}\ \emph {et~al.}(2006)\citenamefont {Shi},
  \citenamefont {Duan},\ and\ \citenamefont {Vidal}}]{Shi2006}%
  \BibitemOpen
  \bibfield  {author} {\bibinfo {author} {\bibfnamefont {Y.-Y.}\ \bibnamefont
  {Shi}}, \bibinfo {author} {\bibfnamefont {L.-M.}\ \bibnamefont {Duan}}, \
  and\ \bibinfo {author} {\bibfnamefont {G.}~\bibnamefont {Vidal}},\ }\href
  {\doibase 10.1103/PhysRevA.74.022320} {\bibfield  {journal} {\bibinfo
  {journal} {Phys. Rev. A}\ }\textbf {\bibinfo {volume} {74}},\ \bibinfo
  {pages} {022320} (\bibinfo {year} {2006})}\BibitemShut {NoStop}%
\bibitem [{\citenamefont {Vidal}(2008)}]{Vidal2008}%
  \BibitemOpen
  \bibfield  {author} {\bibinfo {author} {\bibfnamefont {G.}~\bibnamefont
  {Vidal}},\ }\href {\doibase 10.1103/PhysRevLett.101.110501} {\bibfield
  {journal} {\bibinfo  {journal} {Phys. Rev. Lett.}\ }\textbf {\bibinfo
  {volume} {101}},\ \bibinfo {pages} {110501} (\bibinfo {year}
  {2008})}\BibitemShut {NoStop}%
\bibitem [{\citenamefont {Evenbly}\ and\ \citenamefont
  {Vidal}(2014)}]{Evenbly2014}%
  \BibitemOpen
  \bibfield  {author} {\bibinfo {author} {\bibfnamefont {G.}~\bibnamefont
  {Evenbly}}\ and\ \bibinfo {author} {\bibfnamefont {G.}~\bibnamefont
  {Vidal}},\ }\href {\doibase 10.1103/PhysRevLett.112.240502} {\bibfield
  {journal} {\bibinfo  {journal} {Phys. Rev. Lett.}\ }\textbf {\bibinfo
  {volume} {112}},\ \bibinfo {pages} {240502} (\bibinfo {year}
  {2014})}\BibitemShut {NoStop}%
\bibitem [{\citenamefont {Felser}\ \emph {et~al.}(2021)\citenamefont {Felser},
  \citenamefont {Notarnicola},\ and\ \citenamefont
  {Montangero}}]{PhysRevLett.126.170603}%
  \BibitemOpen
  \bibfield  {author} {\bibinfo {author} {\bibfnamefont {T.}~\bibnamefont
  {Felser}}, \bibinfo {author} {\bibfnamefont {S.}~\bibnamefont {Notarnicola}},
  \ and\ \bibinfo {author} {\bibfnamefont {S.}~\bibnamefont {Montangero}},\
  }\href {\doibase 10.1103/PhysRevLett.126.170603} {\bibfield  {journal}
  {\bibinfo  {journal} {Phys. Rev. Lett.}\ }\textbf {\bibinfo {volume} {126}},\
  \bibinfo {pages} {170603} (\bibinfo {year} {2021})}\BibitemShut {NoStop}%
\bibitem [{\citenamefont {Adam}(1997)}]{Adam1997}%
  \BibitemOpen
  \bibfield  {author} {\bibinfo {author} {\bibfnamefont {C.}~\bibnamefont
  {Adam}},\ }\href {\doibase https://doi.org/10.1006/aphy.1997.5697} {\bibfield
   {journal} {\bibinfo  {journal} {Annals of Physics}\ }\textbf {\bibinfo
  {volume} {259}},\ \bibinfo {pages} {1 } (\bibinfo {year} {1997})}\BibitemShut
  {NoStop}%
\end{thebibliography}%
\end{document}